\documentclass[journal]{IEEEtran}
\usepackage{tikz}
\usetikzlibrary{shapes.geometric, arrows, positioning, fit, backgrounds, calc}
\usepackage{adjustbox}
\usepackage{amsfonts}
\usepackage{amsmath}
\usepackage{array}
\usepackage{bm}
\usepackage{booktabs}
\usepackage{cite}
\usepackage{color}
\usepackage{float}
\usepackage{graphicx}
\usepackage{indentfirst}
\usepackage{longtable}
\usepackage{multirow}
\usepackage{pifont}
\usepackage{subfig}
\usepackage{tabularx}
\usepackage{threeparttable}
\usepackage{wasysym}
\usepackage{yhmath}
\usepackage{newclude} % No a new page when using \include*{}
\usepackage{caption} %in IEEEtrans, {\scshape} or {\sc} is the font of a table title; {singlelinecheck=off} makes a left-aligned title for a figure
\usepackage{hyperref}
\usepackage{balance} 
\usepackage{amssymb}

\title{AI Signal Processing Paradigm for Movable Antenna: From Spatial Position Optimization to Electromagnetic Reconfigurability}

\author{%
Yining Li, 
Ziwei Wan, 
Chongjia Sun, 
Kaijun Feng, 
Keke Ying, 
Wenyan Ma,
Lipeng Zhu,
Xiaodan Shao,
Weidong Mei,
Wenqian Shen,
Zhenyu Xiao,
Zhen Gao, and
Rui Zhang
}

\begin{document}
\maketitle

%%%%%% Abstract %%%%%%
\begin{abstract}  
As 6G wireless communication systems evolve toward intelligence, high reconfigurability, and space-air-ground integration \cite{liu2025toward, liu2024near}, the limitations of traditional fixed antenna (TFA) have become increasingly prominent. As a remedy, spatially movable antenna (SMA) and electromagnetically reconfigurable antenna (ERA) have respectively emerged as key technologies to break through this bottleneck. SMA activates spatial degree of freedom (DoF) by dynamically adjusting antenna positions, ERA regulates radiation characteristics using tunable metamaterials, thereby introducing DoF in the electromagnetic domain. However, the ``spatial-electromagnetic dual reconfiguration" paradigm formed by their integration poses severe challenges of high-dimensional hybrid optimization to signal processing. To address this issue, we integrate the spatial optimization of SMA and the electromagnetic reconfiguration of ERA, propose a unified modeling framework termed movable and reconfigurable antenna (MARA) and investigate the channel modeling and spectral efficiency (SE) optimization for MARA. Besides, we systematically review artificial intelligence (AI)-based solutions, focusing on analyzing the advantages of AI over traditional algorithms in solving high-dimensional non-convex optimization problems. This paper fills the gap in existing literature regarding the lack of a comprehensive review on the AI-driven signal processing paradigm under spatial-electromagnetic dual reconfiguration and provides theoretical guidance for the design and optimization of 6G wireless systems with advanced MARA.

\end{abstract}

\begin{IEEEkeywords}
Spatially movable antenna (SMA), electromagnetically reconfigurable antenna (ERA), movable and reconfigurable antenna (MARA), artificial intelligence (AI), spectral efficiency (SE).
\end{IEEEkeywords}

%%%%%%%%%%%% Main Body%%%%%%%%%%%%

\section{Introduction}\label{S1}
\subsection{Motivation}\label{S1.1}
\begin{figure}[htbp]
\centering
    \includegraphics[width=1\linewidth]{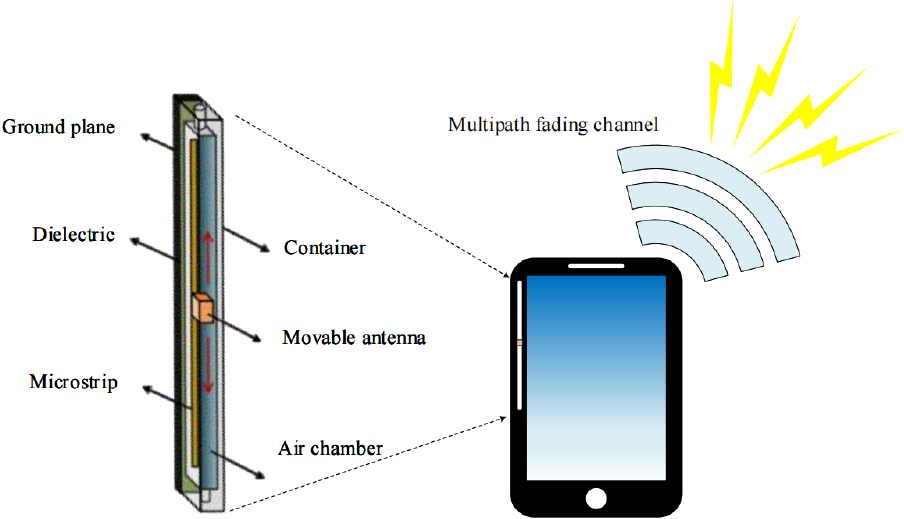}
\caption{Schematic of a SMA and its operation in a multipath channel.}
\label{Fig.gmamanual}
\end{figure}

Contemporary wireless communication systems are advancing into a new era characterized by intelligence and high reconfigurability. Against this backdrop, spatially movable antenna (SMA) technology has attracted extensive attention as an emerging architecture, as shown in Fig. \ref{Fig.gmamanual}. Unlike traditional fixed antenna (TFA) that can only passively endure the adverse impacts of random channel fading, SMA enables spatial dimension reconstruction of wireless channels by allowing flexible movement of antennas at either the transmitter or receiver \cite{1}. This paradigm shift from fixed to movable antennas is deemed to bring transformative opportunities for next-generation networks such as 6G. To support the envisioned massive connectivity and ultra-reliable low-latency communications (URLLC) in 6G \cite{gao2024compressive_iot, ding2023next, ke2021massive, gao2020compressive}, innovative antenna technologies are indispensable. Studies have demonstrated that SMA systems can significantly outperform conventional TFA systems even with the same or fewer number of antennas and radio frequency (RF) chains, with significant performance advantages in terms of received signal power enhancement, interference suppression, flexible beam adjustment, and multiplexing gain \cite{2,3}, which is a natural evolution from the traditional large-scale MIMO systems \cite{gao2014structured}. Essentially, SMA endows wireless systems with previously underutilized spatial degree of freedom (DoF). By dynamically adjusting antenna positions or orientations, the system can adapt to changes in the channel environment, avoid deep channel fading, and thereby achieve superior transmission conditions. Particularly in scenarios with a limited number of antennas, SMA can fully exploit spatial diversity and beamforming gain, thus achieving a better balance between spectral efficiency (SE) and hardware costs \cite{4}. Notably, the advantages of movable antennas are not confined to the communication domain. Recent research has also revealed their significant potential in wireless sensing \cite{buchong1,buchong2}. Optimizing the spatial arrangement of antenna arrays can substantially improve sensing accuracy, providing novel insights for future integrated sensing and communications (ISAC) scenarios \cite{5, gao2023integrated, liao2025integrated}.

% \begin{figure}[!t]
% 	%\vspace*{-5mm}
% 	\begin{center}
% 		\includegraphics[width = 1\columnwidth]{fig/movable antenna.pdf}
% 	\end{center}
% 	%\vspace*{-5mm}
% 	\captionsetup{font={footnotesize}, singlelinecheck=off, name={Fig.}, labelsep=period}
% 	\caption{Schematic of a SMA and its operation in a multipath channel.}
% 	\label{Fig.gmamanual} 
% 	\vspace*{-5mm}
% \end{figure}

\begin{figure}[htbp]
\centering
    \includegraphics[width=0.9\linewidth]{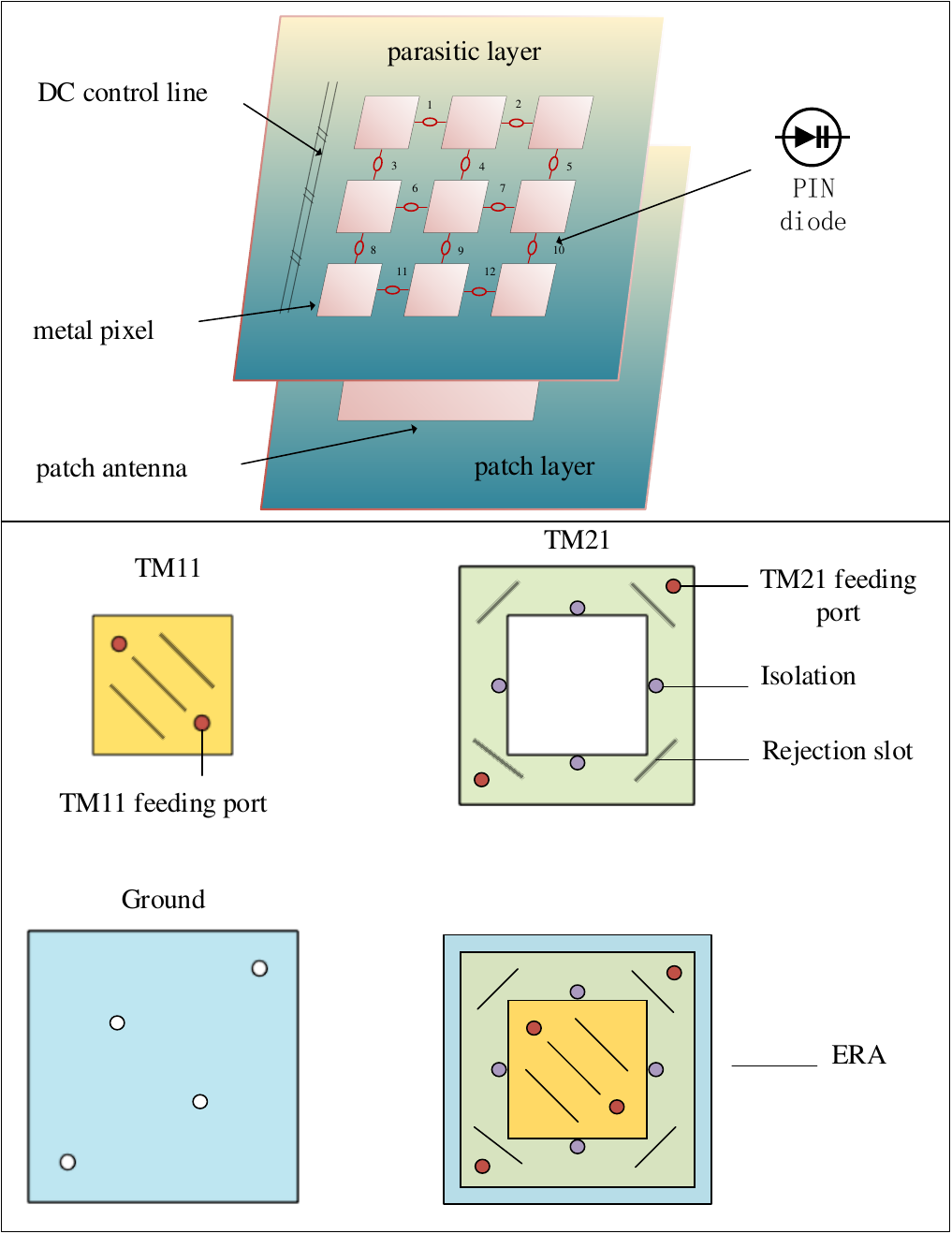}
\caption{Two primary implementation methodologies for ERA.}
\label{Fig.eramanual}
\end{figure}

\begin{figure*}[!t]
	%\vspace*{-5mm}
	\begin{center}
		\includegraphics[width = 2\columnwidth]{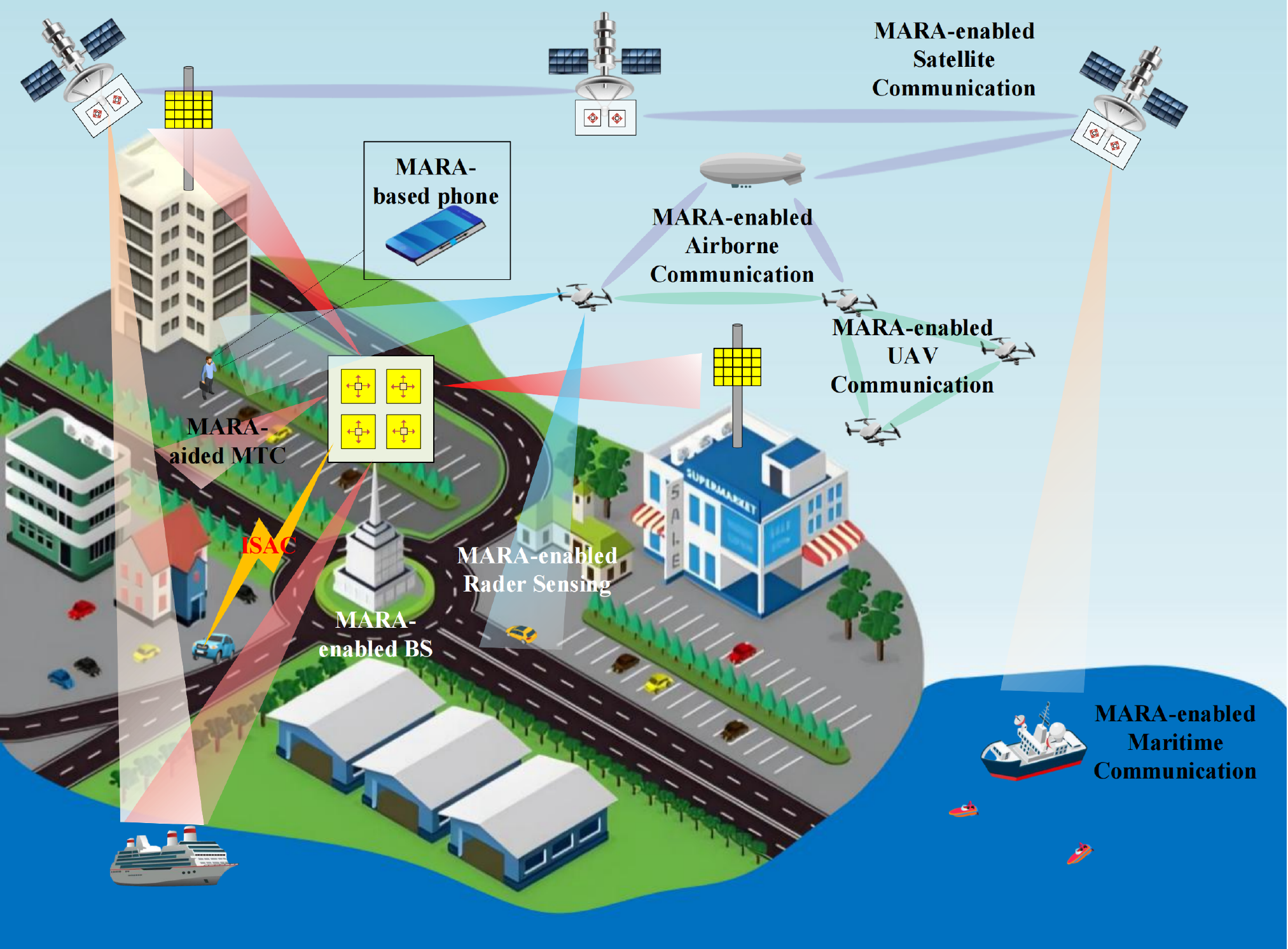}
	\end{center}
	%\vspace*{-5mm}
	% \captionsetup{font={footnotesize}, singlelinecheck=off, name={Figure}, labelsep=period}
	\caption{Application scenarios of MARA-aided wireless networks.}
	\label{Fig.scenario} 
	\vspace*{-5mm}
\end{figure*}

However, spatial movement is insufficient to completely break through the limitations of wireless propagation. The rise of electromagnetic reconfigurable technology has opened another avenue for antenna performance optimization. Once traditional antennas are designed and manufactured, their electromagnetic characteristics remain essentially fixed during operation. In contrast, electromagnetically reconfigurable antenna (ERA) can dynamically regulate its electromagnetic radiation property. Fig. \ref{Fig.eramanual} illustrates two primary implementation methodologies for ERA, including pixel antenna and patch antenna. On this basis, the newly proposed reconfigurable massive MIMO (RmMIMO) architecture adopts reconfigurable pixelated antenna elements, enabling the radiation patterns of each antenna to be adjusted on demand, thereby introducing additional DoF for information transmission in the electromagnetic domain \cite{7}. Note that the concept of ERA has been implicitly utilized in communication scenarios, such as reconfigurable intelligent surfaces (RIS) \cite{wan2021} and holographic surfaces \cite{9}. By flexibly performing amplitude and phase reconstruction of signals at the electromagnetic level, such technologies are equivalent to partially controlling the wireless channel, and can significantly improve system capacity even with limited antenna apertures. 
Given the respect advancement of SMA and ERA, it is foreseeable that combining spatial movement and electromagnetic reconfigurability will endow antenna systems with unprecedented adaptability. On the one hand, antenna positions can be changed to obtain spatial diversity gain. On the other hand, once the antennas have located on the designated positions, their electromagnetic characteristics can be adjusted to shape ideal beams and radiation modes, thus further improving communication quality. The synergy between SMA and ERA is expected to break through the bottlenecks of traditional wireless systems in terms of array gain and SE. We refer to this dual reconfigurable paradigm as movable and reconfigurable antenna (MARA). As shown in Fig. \ref{Fig.scenario}, MARA can be applied to various communication and sensing systems such as MARA-enabled rader sensing, MARA-enabled UAV communication \cite{gao2022data}, MARA-enabled satellite communication \cite{ying2023quasi, li2025active}, etc., enhancing their performance.

Nonetheless, the introduction of MARA with dual reconfigurable DoFs poses unprecedented challenges to signal processing. First, the continuous adjustability of antenna positions implies an infinite number of candidate positions in theory, rendering the extremely high-dimensional non-convex optimization. Particularly in scenarios involving cooperative movement of multiple antennas/subarrays within large regions or multi-user communication, position optimization must account for complex coupling effects, and the solution difficulty far exceeds that of traditional finite-dimensional problems \cite{10,buchong3,buchong4}. Second, electromagnetic reconfiguration introduces a large number of adjustable parameters, resulting in an extremely vast parameter combination space. By jointly considering the spatial layout and electromagnetic characteristics of antennas, solving for the optimal system configuration becomes exceptionally challenging, where high-dimensional hybrid optimization problems are beyond the capability of classical algorithms. For instance, the joint design of digital precoding and analog precoding in RmMIMO requires iterative search over a continuous manifold space, making it difficult to guarantee a globally optimal solution \cite{10}.

Faced with the aforementioned dilemmas, artificial intelligence (AI) has brought new solutions to complex communication optimization problems. Data-driven machine learning approaches, especially for deep learning (DL) and reinforcement learning (RL), can automatically extract features from large-scale samples and approximately learn optimization strategies, enabling the discovery of near-optimal solutions in high-dimensional non-convex spaces \cite{12}. Notably, existing studies have begun to explore the application of AI in the design of SMA and ERA systems. For example, using deep neural networks (DNN) to predict the channel gain of potential antenna ports, or leveraging RL to intelligently search for optimal antenna movement trajectories \cite{13}. The powerful self-learning and generalization capabilities of AI enable it to continuously track dynamic channels and real-time optimize system parameters, thereby fully unleashing the potential of movement optimization and electromagnetic reconfiguration. It can be expected that in the era of MARA, AI is no longer an optional tool, but a key enabler to resolve system complexity and drive significant performance leaps.

% \begin{figure*}[!t]
% 	%\vspace*{-5mm}
% 	\begin{center}
% 		\includegraphics[width = 2\columnwidth]{fig/scenario.pdf}
% 	\end{center}
% 	%\vspace*{-5mm}
% 	\captionsetup{font={footnotesize}, singlelinecheck=off, name={Fig.}, labelsep=period}
% 	\caption{Application scenarios of MARA-aided wireless networks.}
% 	\label{Fig.scenario} 
% 	\vspace*{-5mm}
% \end{figure*}

\subsection{Comparisons and Key Contributions}\label{S1.2}

While this field holds great promise, current research remains in its infancy, and existing literature has not yet provided a comprehensive review of AI-driven signal processing paradigms under spatial-electromagnetic dual reconfiguration. Recently, several representative works have touched upon relevant subfields, each with distinct focuses and limitations.

For SMA, Zhu et al. systematically outlined the evolution of SMA technology and its performance gains in communication and sensing in their tutorial, emphasizing how the transition from TFA to SMA empowers more flexible and efficient networks \cite{11}. In \cite{zhangrui1}, Shao et al. further introduced the concept of six-dimensional movable antenna (6DMA) which exploits both antenna position and rotation adjustment for enhancing wireless communication and sensing performance. However, these tutorials primarily focus on mechanical antenna movement optimization, devoting limited attention to the design of ERAs.

For electromagnetic reconfiguration, Ying et al. proposed the architecture of RmMIMO and, in subsequent studies, delved into precoding design and channel estimation schemes based on electromagnetic-domain CSI \cite{14}. These works demonstrate the significant value of reconfigurable antennas in improving system capacity with limited apertures. However, they primarily focus on the optimization of antenna electromagnetic parameters, do not address the dynamic adjustment of antenna spatial positions, and overlook the potential of integrating AI to simplify complex algorithm design.

Overall, most existing studies are confined to individual subfields (e.g., SMA/6DMA, fluid antennas, or electromagnetic reconfiguration), lacking a holistic perspective and integrated analysis of the entire landscape of generally movable antenna signal processing.
Against this backdrop, the unique contributions of this paper lie in: 
\begin{itemize}
	\item 
	Perspective of Innovative Integration: For the first time, it integrates and contextualizes the two research threads of spatial optimization for movable antennas and electromagnetic reconfiguration, systematically summarizing the development status and key principles of the dual-reconfigurable antenna system.
	\item
	AI-Empowered Core Thread: It provides a comprehensive review of the application progress of AI technologies in this field, discussing how the methods such as DL and RL are applied to solve challenges including antenna position optimization, CSI acquisition, beamforming, and adaptive beam control. It also compares their advantages and disadvantages with those of traditional algorithms.
	\item
	Direction for Future Research: By comparing the state-of-the-art studies, we identify the shortcomings of current research in both theoretical and practical aspects, such as the lack of experimental data for movable antennas under real-world channels, limited interpretability and generalization ability of AI models, and the absence of a unified framework for spatial-electromagnetic joint optimization. Based on these insights, we prospect potential future breakthrough directions accordingly.
\end{itemize}

\subsection{Organization}\label{S1.3}	

The remainder of this paper is organized as follows. Section II establishes the fundamentals of MARA, it classifies SMA, ERA, and MARA, clarifies their principles and trade-offs, and builds a unified mathematical framework for their channel modeling and SE optimization. Section III details AI-driven solutions for MARA, focusing on core scenarios like CSI estimation, beamforming optimization, and ISAC. Section IV outlines promising future directions and associated challenges. Finally, this paper is concluded in Section V.

Fig. \ref{Fig.organization} presents a flowchart that illustrates the overall organization of this review paper. The flowchart depicts the logical flow from the introduction and motivation (Section I), through the fundamentals of MARA including classification and unified modeling framework (Section II), to the comprehensive review of AI-empowered approaches for channel estimation, beamforming, and ISAC (Section III), and finally to future research directions (Section IV) and conclusions (Section V).

\begin{figure}[htbp]
\centering
\begin{tikzpicture}[
    node distance=0.4cm,
    startstop/.style={rectangle, rounded corners, minimum width=2.6cm, minimum height=0.55cm, text centered, draw=black, fill=blue!20, font=\scriptsize},
    process/.style={rectangle, minimum width=2.6cm, minimum height=0.55cm, text centered, draw=black, fill=orange!20, font=\scriptsize},
    subbox/.style={rectangle, minimum width=1.5cm, minimum height=0.4cm, text centered, draw=black, fill=green!15, font=\tiny},
    typebox/.style={rectangle, minimum width=1.3cm, minimum height=0.35cm, text centered, draw=black, fill=yellow!20, font=\tiny},
    arrow/.style={thick,->,>=stealth}
]

% Section I
\node (sec1) [startstop] {Section I: Introduction};

% Section II
\node (sec2) [process, below=0.5cm of sec1] {Section II: Fundamentals};

% Section II linear flow: Classification -> MA Types (horizontal) -> Unified Modeling -> Key Challenges
\node (sec2a) [subbox, below=0.35cm of sec2] {Classification};

% MA types - horizontal arrangement with wider spacing
\node (sma) [typebox, below=0.35cm of sec2a, xshift=-1.8cm] {SMA};
\node (era) [typebox, below=0.35cm of sec2a] {ERA};
\node (mara) [typebox, below=0.35cm of sec2a, xshift=1.8cm] {MARA};

\node (sec2b) [subbox, below=0.7cm of era] {Unified Modeling};
\node (sec2c) [subbox, below=0.35cm of sec2b] {Key Challenges};

% Section III
\node (sec3) [process, below=0.5cm of sec2c] {Section III: AI-Empowered Approaches};
\node (sec3a) [subbox, below=0.35cm of sec3, xshift=-1.6cm] {Channel Est.};
\node (sec3b) [subbox, below=0.35cm of sec3] {Beamforming};
\node (sec3c) [subbox, below=0.35cm of sec3, xshift=1.6cm] {ISAC};

% Convergence point for Section III subsections
\coordinate (conv3) at ($(sec3b.south)+(0,-0.5cm)$);

% Section IV with 3 sub-directions - evenly spaced
\node (sec4) [process, below=1.3cm of sec3] {Section IV: Future Directions};
\node (sec4b) [subbox, below=0.35cm of sec4, xshift=-1.8cm, minimum width=1.2cm] {Novel Waveform};
\node (sec4c) [subbox, below=0.35cm of sec4, minimum width=1.2cm] {Embodied Intel.};
\node (sec4d) [subbox, below=0.35cm of sec4, xshift=1.8cm, minimum width=1.2cm] {Multi-stage BF};

% Convergence point for Section IV subsections
\coordinate (conv4) at ($(sec4.south)+(0,-1.1cm)$);

% Section V
\node (sec5) [startstop, below=1.5cm of sec4] {Section V: Conclusions};

% Main arrows
\draw [arrow] (sec1) -- (sec2);
\draw [arrow] (sec2) -- (sec2a);

% Classification to MA types (horizontal)
\draw [arrow] (sec2a) -- (sma);
\draw [arrow] (sec2a) -- (era);
\draw [arrow] (sec2a) -- (mara);

% MA types converge to Unified Modeling
\draw [arrow] (sma) -- (sec2b);
\draw [arrow] (era) -- (sec2b);
\draw [arrow] (mara) -- (sec2b);

% Unified Modeling to Key Challenges
\draw [arrow] (sec2b) -- (sec2c);

% Key Challenges to Section III
\draw [arrow] (sec2c) -- (sec3);

\draw [arrow] (sec3) -- (sec3a);
\draw [arrow] (sec3) -- (sec3b);
\draw [arrow] (sec3) -- (sec3c);

% Three subsections converge to a point, then to Section IV
\draw [thick] (sec3a.south) -- (conv3);
\draw [thick] (sec3b.south) -- (conv3);
\draw [thick] (sec3c.south) -- (conv3);
\draw [arrow] (conv3) -- (sec4);

\draw [arrow] (sec4) -- (sec4b);
\draw [arrow] (sec4) -- (sec4c);
\draw [arrow] (sec4) -- (sec4d);

% Three directions converge to a point, then to Section V
\draw [thick] (sec4b.south) -- (conv4);
\draw [thick] (sec4c.south) -- (conv4);
\draw [thick] (sec4d.south) -- (conv4);
\draw [arrow] (conv4) -- (sec5);

\end{tikzpicture}
\caption{Flowchart of the organization of this paper.}
\label{Fig.organization}
\end{figure}

\section{Movable Antenna and Beyond: Fundamentals and Principles}\label{S2}

This section presents a unified mathematical framework that consolidates the channel modeling and optimization formulations for SMA, ERA, and the proposed MARA paradigm. By establishing consistent notations across different antenna architectures, we aim to facilitate direct comparison and provide a common foundation for the AI-based methods reviewed in Section III.

\subsection{Classification}\label{S2.1}

There are multiple definitions of a movable antenna. In the narrow sense, an antenna that can be repositioned in physical space is considered movable; in the broad sense, an antenna is regarded as movable if any of its intrinsic properties can be altered, for example, by reconfiguring its radiation pattern \cite{11}. This section categorizes movable antennas into two types: 1) SMA, which achieves mobility through spatial repositioning, and 2) ERA, which achieves mobility by dynamically modifying electromagnetic characteristics.

\subsubsection{SMA}\label{S2.2.1}
\begin{figure}
	%\vspace*{-5mm}
	\begin{center}
		\includegraphics[width = 1\columnwidth]{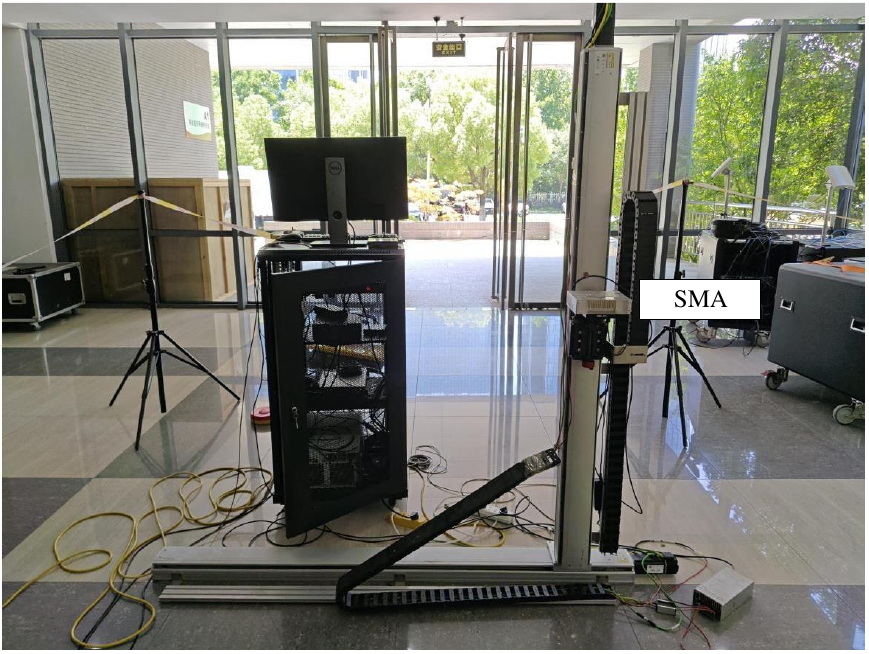}
	\end{center}
	%\vspace*{-5mm}
	% \captionsetup{font={footnotesize}, singlelinecheck=off, name={Fig.}, labelsep=period}
	\caption{Mechanical slide-based SMA communication prototype developed by Southeast University in \cite{fig1}.}
	\label{Fig.gmareal} 
	\vspace*{-5mm}
\end{figure}

Such antennas adjust their position or orientation through macroscopic physical movement, with their core characteristic being the reconfiguration of spatial location. The implementation of SMA involves two scales: element-level movement and array-level movement. At the element level, a typical approach employs external mechanical structures such as motors, gears, or micro-electro-mechanical systems (MEMS) devices to drive the displacement or rotation of individual antenna elements \cite{1,15}. These actuators convert control signals into precise mechanical motions, enabling the antenna to be positioned at an optimal location or orientation within the transmission/reception region. Fig. \ref{Fig.gmareal} shows an example of an SMA.

Furthermore, fluidic antennas, such as those based on liquid metals, represent another important form of SMA. By utilizing the fluidity of liquid materials, liquid metals can be driven to flow through microchannels within an encapsulated structure via devices such as syringes or micropumps, thereby modifying the antenna's shape and position \cite{16.1,16.2,16.3,16.4,16.5}. This liquid-based reconfigurable approach allows the radiating elements of the antenna to slide continuously in space or switch among predefined positions, adapting to variations in the communication channel.

It should be noted that although mechanically movable antennas can enhance wireless communication and sensing performance, they also face several challenges. First, the hardware complexity and maintenance costs are relatively high. For instance, motors and transmission mechanisms increase system weight and are subject to mechanical wear. Second, the response speed is generally limited by the inertia of mechanical motion, with response times ranging from milliseconds to seconds \cite{16.1,16.5}. Third, variations in electromagnetic coupling may occur among multiple movable antenna elements during movement, necessitating careful design to mitigate adverse inter-antenna effects.

Nevertheless, owing to their extensive physical movement range and high degrees of freedom, SMAs can achieve substantial improvements in channel gain. For example, in narrowband slow-fading scenarios, adjusting the antenna position can significantly enhance the received signal strength or suppress interference \cite{17.1,17.2}. Existing studies have shown that even within a very limited space, if a single fluidic antenna can switch among a sufficient number of positions, its outage performance can surpass that of a multi-antenna maximum ratio combining (MRC) receiver \cite{3,18}. As the antenna movement region expands further, significant performance gains over TFA systems can be realized \cite{buchong3,buchong4}. Therefore, as a novel antenna architecture that fully exploits spatial diversity, SMAs demonstrate considerable potential for future networks such as 6G.

\subsubsection{ERA}\label{S2.2.1}

\begin{figure}[!t]
	%\vspace*{-5mm}
	\begin{center}
		\includegraphics[width = 1\columnwidth]{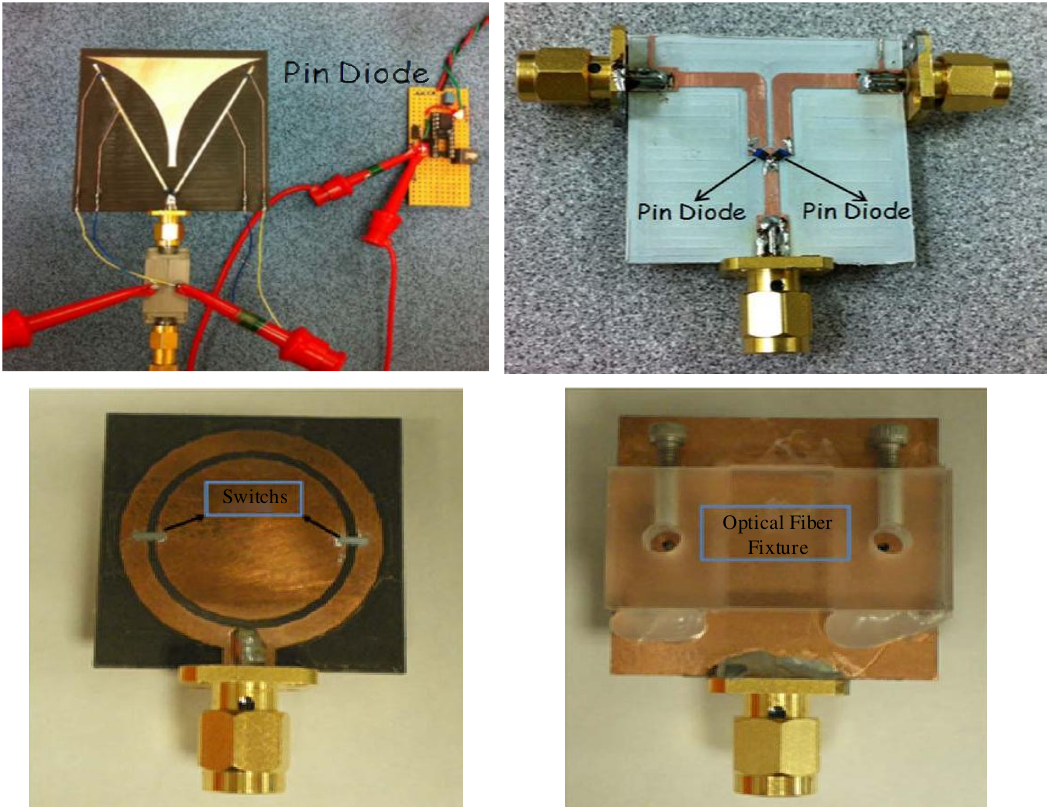}
	\end{center}
	%\vspace*{-5mm}
	% \captionsetup{font={footnotesize}, singlelinecheck=off, name={Fig.}, labelsep=period}
	\caption{Some practical schematic diagrams of ERA taken from \cite{7}.}
	\label{Fig.erareal} 
	\vspace*{-5mm}
\end{figure}

Such antennas change the equivalent position or radiation pattern of the antenna through the reconstruction of microscopic electromagnetic properties, rather than relying on macroscopic mechanical displacement. The core of ERA's operation is to introduce a new adjustable control dimension in the electromagnetic domain, dynamically adjust the antenna's radiation parameters such as radiation pattern and polarization, thereby equivalently changing the antenna's observation method for the channel. Fig. \ref{Fig.erareal} presents a series of examples of ERA.

A typical example is the multimode patch antenna, by exciting different eigenmodes in the antenna structure (e.g., the TM11 mode and TM21 mode in the stacked circular patch antenna), the phase center of the antenna can be shifted from its physical center, realizing the reconstruction of the radiation pattern \cite{19.1,19.2}. Such mode switching allows the antenna to achieve an effect similar to the position change of a movable antenna without moving the physical position of the antenna itself.

Another example is the pixelated antenna, which is composed of a large number of tiny radiating unit pixels. By controlling the on-off states of each pixel via electronic switches, the shape and effective radiation aperture of the antenna can be dynamically changed. Turning on the pixels in a specific area is equivalent to generating a radiating unit at that position, while turning them off is equivalent to hiding it. Therefore, the pixelated antenna can instantly appear or disappear at different positions in space, which is equivalent to realizing the electronic switching of the antenna position \cite{20}.

A prominent advantage of ERA is its fast response speed, since it relies entirely on RF devices such as electrically controlled switches or phase shifters, its reconfiguration rate can reach the nanosecond to microsecond level, far faster than the mechanically moving SMA \cite{21}. This enables ERA to adjust itself at a high frequency to track fast-changing channels. At the same time, ERA does not require support from large mechanical structures, avoiding wear and tear as well as space constraints, making it highly suitable for compact wireless systems with high requirements for volume and reliability.

By endowing each antenna with adjustable radiation patterns, ERA introduces additional DoF in electromagnetic radiation modes that traditional fixed antennas do not possess, and is expected to further enhance information transmission capability. For instance, reconfigurable pixel antenna technology provides each antenna with adjustable radiation patterns, allowing the system to add a new degree of freedom in the electromagnetic domain for information transmission and increase the upper limit of capacity under a given array size. Preliminary studies have shown that after introducing ERA, the SE and energy efficiency of the system can be significantly improved through the appropriate design of multi-level beamforming schemes \cite{10,14,21}.

\begin{figure*}[!t]
	%\vspace*{-5mm}
	\begin{center}
		\includegraphics[width = 1.8\columnwidth]{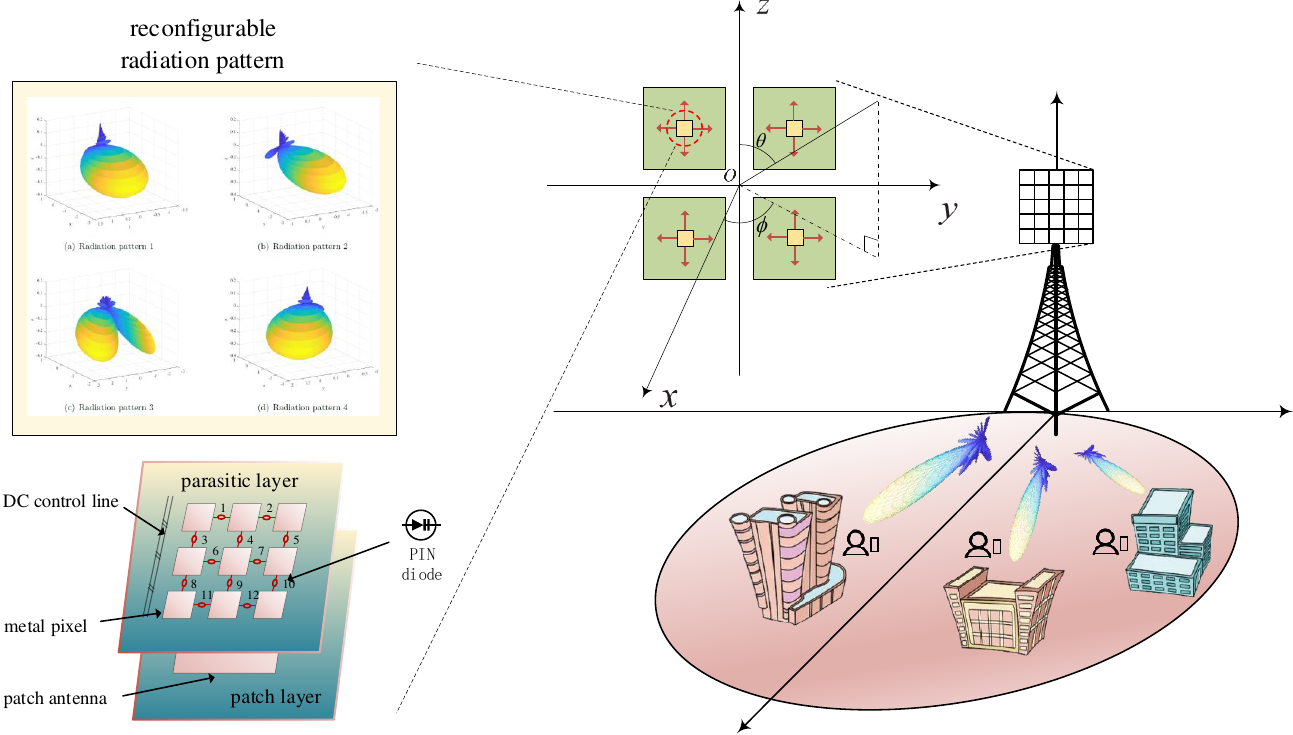}
	\end{center}
	%\vspace*{-5mm}
	% \captionsetup{font={footnotesize}, singlelinecheck=off, name={Fig.}, labelsep=period}
	\caption{MARA-based multi-user downlink transmission.}
	\label{Fig.system} 
	\vspace*{-5mm}
\end{figure*}

However, ERA also has its limitations. First, the hardware cost required to realize electromagnetic reconstruction is high, and the control circuits are complex. Especially for large-scale reconfigurable arrays, achieving independent control of each unit requires substantial overhead. Second, compared with real spatial displacement, the phase center shift range and radiation pattern coverage range that ERA can achieve are limited, and its equivalent movement capability is restricted by the antenna structure and operating mode to a certain extent.

Nevertheless, by expanding new adjustment methods in the electromagnetic domain, ERA provides a novel approach for improving the performance of ultra-large-scale antenna systems. For example, after adopting reconfigurable antennas, base stations (BSs) can still partially control the wireless channel by changing the antenna radiation pattern under the condition of a given limited physical size, achieving performance gains that are unattainable with traditional fixed arrays.

\subsubsection{Summary}\label{S2.2.1}

Overall, SMA and ERA represent two orthogonal directions of antenna reconfigurability paradigms in the spatial domain and electromagnetic domain respectively. The former directly changes the spatial position where the antenna is located, while the latter achieves an equivalent effect by altering the electromagnetic radiation characteristics of the antenna. Each of the two has its own advantages and disadvantages and complements each other. SMA offers a large physical movement range and intuitive spatial gain, but its response is relatively slow due to the limitations of mechanical structures. ERA has a fast response and does not require large-scale moving components, but is constrained by circuit complexity and mode range. Because of this, in practical applications, the appropriate type can be selected based on requirements, or the two methods can be integrated to leverage the advantages of both aspects.

\subsection{Problem Formulation}\label{S2.2}

Currently, there are numerous studies on SMA and ERA, however, no research has yet explored their inherent deep-seated connections and organically unified the two. This paper innovatively proposes a modeling approach that can unify SMA and ERA. Next, we will proceed from simplicity to complexity. We begin by introducing the modeling approach for SMA, followed by that for ERA. Finally, we integrate both within the MARA framework to achieve a unified representation of movable antennas.

To ensure the universality of the proposed scheme, the most complex scenario is considered for all the three aforementioned schemes, i.e., the downlink wideband transmission problem from a massive MIMO (mMIMO) BS to \( U \) single-antenna user equipments (UEs) in a time division duplex (TDD) system, as shown in Fig. \ref{Fig.system}. Regardless of the antenna type, the BS employs a fully digital architecture. This array consists of \( M \) antenna elements. Meanwhile, we adopt orthogonal frequency division multiplexing (OFDM) to combat frequency-selective fading.

In addition, to avoid redundant calculation of simplification processes, the single antenna at the UE side is set as TFA which is neither spatially movable nor electromagnetically reconfigurable. The only difference between different schemes lies in the variation of antenna properties at the BS side.

\subsubsection{Channel Modeling and Optimization Problem of SMA}\label{S2.2.1}

In this section, we set the BS antennas as spatially movable but non-electromagnetically reconfigurable antennas. For the \( u \)-th UE (where \( 1 \leq u \leq U \)), the received signal at the \( g \)-th subcarrier is expressed by the following formula
\begin{equation}
		\label{yug}
		y_{u,g} = \boldsymbol{h}_{u,g}^H \boldsymbol{W}_g \boldsymbol{s}_g + n_{u,g}, \quad 1 \leq g \leq G,
\end{equation}
where \( G \) stands for the total count of subcarriers, \( \boldsymbol{h}_{u,g}^H = [h_{u,1,g}^{\rm SMA}, h_{u,2,g}^{\rm SMA}, \dots, h_{u,M,g}^{\rm SMA}] \in \mathbb{C}^{1 \times M} \) signifies the downlink channel from the BS to the \( u \)-th UE, \( \boldsymbol{W}_g \in \mathbb{C}^{M \times U} \) serves as the digital precoder, \( \boldsymbol{s}_g \in \mathbb{C}^U \) indicates the transmitted signal over the \( g \)-th subcarrier, and \( n_{u,g} \) refers to the complex additive white Gaussian noise (AWGN) at the UE.

For the downlink channel between the \( m \)-th antenna and the \( u \)-th UE over the \( g \)-th subcarrier, it may be written as follows
\begin{equation}
		\label{humgbasic}
		\begin{aligned} 
		h_{u, m, g}^{\rm SMA} = \sum_{i=1}^{L_{u}} & \tilde{x}_{i, u}  
		  e^{-j \frac{2 \pi}{\lambda}\left(\boldsymbol{k}_{\text{tx}, i, u}^{T} \boldsymbol{p}_{m} + \boldsymbol{k}_{\text{rx}, i, u}^{T} \boldsymbol{q}_{u}\right)} \cdot e^{-j 2 \pi \tau_{i, u} f_{g}},
		\end{aligned}
\end{equation}
where \( \tilde{x}_{i, u} \) represents the complex channel coefficient for the \( i \)-th propagation path associated with the \( u \)-th UE, \( \lambda \) stands for the carrier wavelength, \( \boldsymbol{k}_{\text{tx}, i, u} \) and \( \boldsymbol{k}_{\text{rx}, i, u} \) are defined as the wave vectors for the \( i \)-th path at the transmitting and receiving ends, respectively, \( \boldsymbol{p}_{m} \) specifies the position vector of the \( m \)-th transmitting antenna, \( \boldsymbol{q}_{u} \) gives the position vector of the antenna for the \( u \)-th UE, and \( f_g \) indicates the operating frequency of the \( g \)-th subcarrier. The BS antennas are permitted to adjust their positions within specified boundaries, subject to the following positional constraints: \( \mathcal{P}_m = \left\{ \boldsymbol{p} \in \mathbb{R}^3: \| \boldsymbol{p} - \boldsymbol{p}_m \|_2 < \frac{d}{2} \right\} \), where \( d \)  corresponds to the inter-antenna spacing in the initial configuration, characterized by uniform separation between all antenna pairs. Furthermore, \( L_u \), \( \tilde{x}_{i, u} \), \( \tau_{i, u} \), and \( \lambda \) correspond to the multipath component count, complex channel coefficient, propagation delay, and carrier wavelength, respectively.

By including the delay component within the channel gain through the definition \(x_{i,u,g}=\tilde{x}_{i,u}e^{-j2\pi\tau_{i,u}f_{g}}\), the channel representation can be simplified to the following compact formulation
\begin{equation}
		\label{humg}
		h_{u,m,g}^{\rm SMA} =\boldsymbol{A}_u \boldsymbol{\Sigma}_{u,g} \boldsymbol{B}_{u,m} ,
\end{equation}
where \( \boldsymbol{B}_{u,m} = \text{diag}( \boldsymbol{b}_{u,m} )\) and \( \boldsymbol{A}_u = \text{diag}( \boldsymbol{a}_u ) \) are diagonal matrices with \( \boldsymbol{b}_{u,m} = \left[ e^{-j \frac{2\pi}{\lambda} \boldsymbol{k}_{\text{tx},1,u}^T \boldsymbol{p}_m}, \dots, e^{-j \frac{2\pi}{\lambda} \boldsymbol{k}_{\text{tx},L_u,u}^T \boldsymbol{p}_m} \right]^T \) and \( \boldsymbol{a}_u = \left[ e^{-j \frac{2\pi}{\lambda} \boldsymbol{k}_{\text{rx},1,u}^T \boldsymbol{q}_u}, \dots, e^{-j \frac{2\pi}{\lambda} \boldsymbol{k}_{\text{rx},L_u,u}^T \boldsymbol{q}_u} \right]^T \), while \( \boldsymbol{\Sigma}_{u,g} = \text{diag}\left( [x_{1,u,g}, \dots, x_{L_u,u,g}] \right) \) contains the complex channel coefficients.

To focus on exploring the gains brought by SMA, in the subsequent research, we assume that the system can obtain perfect CSI, and we only need to consider the improvement in SE of SMA compared to TFA-based systems. In the aforementioned channel model, the BS antennas can maximize SE by adjusting \( \boldsymbol{B}_{u,m} \) and the corresponding precoding matrix \( \boldsymbol{W}_g \).

According to (\ref{yug}), the total SE \(R_g\) of the system is given by
\begin{equation}
	\label{R1}
	R_{\rm SMA} = \sum_{g=1}^G \sum_{u=1}^U \log_2 \left( 1 + \frac{\left| \boldsymbol{h}_{u,g}^H \boldsymbol{w}_{u,g} \right|^2}{\sum_{u' \neq u}^U \left| \boldsymbol{h}_{u',g}^H \boldsymbol{w}_{u',g} \right|^2 + \sigma_n^2} \right),
\end{equation}
where the digital precoder for the \( u \)-th UE is denoted by \( \boldsymbol{w}_{u,g} \), and the composite precoding matrix across all UEs is given by \( \boldsymbol{W}_g = [\boldsymbol{w}_{1,g}, \boldsymbol{w}_{2,g}, \dots, \boldsymbol{w}_{U,g}] \in \mathbb{C}^{M \times U} \).

Then, we can formulate the SE optimization problem for SMA as follows
\begin{align}
\label{SE1}
	\begin{split}
		\mathop{\max}\limits_{\{\bm{p}_{m}\}_{m=1}^{M},\left\{\bm{W}_{g}\right\}_{g=1}^{G}} R_{\rm SMA} \\    
		\qquad\text{s.t.} \quad\sum_{g=1}^{G}\|\bm{W}_g\|_{\mathrm{F}}^{2}\le P_T,\\
        \quad\bm{p}_{m}\in \mathcal{P}_m, \forall m,
	\end{split}
\end{align}
where \( P_T \) denotes the power constraint of the digital precoder.

% \begin{align}\label{SE3}
% 	\begin{split}
% 		\mathop{\max}\limits_{\{\bm{p}_{m}\}_{m=1}^{M},\left\{\bm{\alpha}_m\}_{m=1}^{M},\{\bm{W}_{g}\right\}_{g=1}^{G}} R_{MARA} \\ 
% 		\qquad\text{s.t.} \quad\|\bm{\alpha}_{m}\|^{2} = 1, \forall m, \\ \qquad\qquad\sum_{g=1}^{G}\|\bm{W}_g\|_{\mathrm{F}}^{2}\le P_T, \\
%         \quad\bm{p}_{m}\in \mathcal{P}_m, \forall m.
% 	\end{split}
% \end{align}

\subsubsection{Channel Modeling and Optimization Problem of ERA}\label{S2.2.2}

In this section, we set the BS antennas as electromagnetically reconfigurable but non-spatially movable antennas. Its basic channel model is consistent with that of SMA. Unlike SMA which explicitly changes antenna positions, ERA influences the channel by adjusting the electromagnetic radiation patterns of antennas. Thus, its channel modeling requires incorporating antenna radiation characteristics into the original channel model. Specifically, in addition to the traditional spatial CSI (sCSI), ERA introduces the concept of electromagnetic-domain CSI (eCSI) to characterize the channel's sensitivity to antenna radiation patterns. An effective modeling approach is to utilize spherical harmonic orthogonal decomposition (SHOD) to expand the far-field radiation pattern of an antenna. Via the spherical harmonic series representation, any antenna radiation pattern can be decomposed into a linear combination of a set of orthogonal radiation modes.

The received signal at the \( g \)-th subcarrier is is consistent with equation (\ref{yug}), where \( \boldsymbol{h}_{u,g}^H = [h_{u,1,g}^{\rm ERA}, h_{u,2,g}^{\rm ERA}, \dots, h_{u,M,g}^{\rm ERA}] \in \mathbb{C}^{1 \times M} \) signifies the downlink channel from the BS to the \( u \)-th UE.
\begin{equation}
		\label{humgbasic}
		\begin{aligned} 
		h_{u, m, g}^{\rm ERA} = \sum_{i=1}^{L_{u}} & \tilde{x}_{i, u} f_{\text{rx}, u}\left(\vartheta_{i, u}, \varphi_{i, u}\right) f_{\text{tx}, m}\left(\theta_{i, u}, \phi_{i, u}\right) \\ 
		& \times e^{-j \frac{2 \pi}{\lambda}\left(\boldsymbol{k}_{\text{tx}, i, u}^{T} \boldsymbol{p}_{m} + \boldsymbol{k}_{\text{rx}, i, u}^{T} \boldsymbol{q}_{u}\right)} \cdot e^{-j 2 \pi \tau_{i, u} f_{g}},
		\end{aligned}
\end{equation}
where \( f_{\text{tx}, m}(\theta_{i, u}, \phi_{i, u}) \) indicates the radiation pattern response of the \( m \)-th transmit antenna element towards the direction \( (\theta_{i, u}, \phi_{i, u}) \), \( f_{\text{rx}, u}(\vartheta_{i, u}, \varphi_{i, u}) \) corresponds to the radiation pattern response of the receiving antenna for the \( u \)-th UE in the direction \( (\vartheta_{i, u}, \varphi_{i, u}) \).
We postulate that an arbitrary radiation pattern can be represented as a linear combination of a set of \( K \) orthonormal basis functions \( \{\omega_k(\theta,\phi)\}_{k=1}^K \), that is,
\begin{equation}
	\label{fdecomposition}
	f(\theta,\phi) = \sum_{k=1}^K \alpha_k \omega_k(\theta,\phi),
\end{equation}
where the orthonormal basis functions adhere to the normalized orthogonality condition:$	\iint \omega_{k}(\theta,\phi) \omega_{k^{\prime}}(\theta,\phi)\sin\theta d\theta d\phi = \left\{
\begin{aligned}
	0 \quad k\neq k^{\prime},\\
	1 \quad k = k^{\prime}.\\
\end{aligned}\right.$, and $\left\{\alpha_{k}\right\}_{k=1}^{K}$ are the weight coefficients. 
The channel representation can be simplified to the following compact formulation
\begin{equation}
		\label{humg}
		h_{u,m,g}^{\rm ERA} = \boldsymbol{f}_{\text{rx},u}^T \boldsymbol{A}_u \boldsymbol{\Sigma}_{u,g} \boldsymbol{\tilde{B}}_{u,m} \boldsymbol{f}_{\text{tx},u,m},
\end{equation}
where \( \boldsymbol{f}_{\text{rx},u} \) and \( \boldsymbol{f}_{\text{tx},u,m} \) are the radiation pattern response vectors at the receiver and transmitter sides, respectively. A key distinction is that the position-related variable \( \boldsymbol{p}_{m} \) in \(\boldsymbol{\tilde{B}}_{u,m}\) is immobile, unlike in \(\boldsymbol{{B}}_{u,m}\). Furthermore, the original radiation pattern gain vector at the transmitter is approximated using a truncated SHOD, that is,
\begin{equation}
	\label{Fdecomposition}
	\boldsymbol{f}_{\text{tx},u,m} = \boldsymbol{\Omega}_u\boldsymbol{\alpha}_m, \quad \forall u,m,
\end{equation}
where \( \boldsymbol{\Omega}_u \) is the basis function matrix with \( [\boldsymbol{\Omega}_u]_{i,k} = \omega_k(\theta_{i,u}, \phi_{i,u}) \), and \( \boldsymbol{\alpha}_m = [\alpha_{1,m}, \dots, \alpha_{K,m}]^T \) represents the electromagnetic-domain precoder for the \( m \)-th antenna. Accordingly, the channel coefficient in (\ref{humg}) can be reformulated as
\begin{equation}
	\label{split}
	h_{u,m,g}^{\rm ERA} = \boldsymbol{f}_{\text{rx},u}^T \boldsymbol{A}_u \boldsymbol{\Sigma}_{u,g} \boldsymbol{\tilde{B}}_{u,m} \boldsymbol{\Omega}_u \boldsymbol{\alpha}_m \stackrel{(a)}{=} \boldsymbol{\tilde{q}}_{u,m,g}^H \boldsymbol{\alpha}_m, 
\end{equation}
where we establish the definition $\bm{\tilde{q}}_{u,m,g}^{H} \triangleq \bm{f}_{{\rm{rx}},u}^{T}\bm{A}_{u}\bm{\mathit{\Sigma}}_{u,g}\bm{\tilde{B}}_{u,m}\bm{\mathit\Omega}_u\in \mathbb{C}^{1\times K}$, and \( \boldsymbol{\alpha}_m \) will be designated as the electromagnetic-domain precoder associated with the \( m \)-th antenna in the subsequent analysis. This methodology effectively decouples the original channel into a product of two components: \( \boldsymbol{\tilde{q}}_{u,m,g} \), which encapsulates the environmental characteristics, and \( \boldsymbol{\alpha}_m \), which characterizes the transmitter's radiation pattern. To clearly differentiate this newly defined variable from the conventional channel representation, \( \boldsymbol{\tilde{q}}_{u,m,g} \) is defined as eCSI. Given that eCSI comprehensively captures the channel properties within the electromagnetic domain, it is anticipated to enhance the performance of precoding designs at the transmitter side.

The primary objective of this research centers on the throughput-maximizing design of \( \boldsymbol{\alpha}_m \). Following the acquisition of the optimized \( \boldsymbol{\alpha}_m \), the corresponding antenna radiation pattern for the \( m \)-th element can be regenerated through equation (\ref{fdecomposition}). Diverging from the TFA approach that depends exclusively on sCSI-based digital precoding, our methodology of eCSI-guided \( \boldsymbol{\alpha}_m \) optimization facilitates significant throughput improvements by effectively harnessing the available degrees of freedom within the electromagnetic domain.

Similar to (\ref{R1}), we can obtain the SE of the system as
\begin{equation}\label{R2}
	R_{\rm ERA}\! = \sum\limits_{g=1}^{G}\sum\limits_{u=1}^{U}\! \log_{2}\! \left(\!\! 1\! +\! \frac{|\bm{\tilde{q}}_{u,g}^{ H}{\bm{\mathit\Lambda}}\bm{w}_{u,g}|^2}{\sum\limits_{u^{\prime}\neq u}^{U}\! |\bm{\tilde{q}}_{u,g}^{H}{\bm{\mathit\Lambda}}\bm{w}_{u^{\prime},g}|^2\! +\! \sigma_n^2}\! \right)\!\! , \!
\end{equation}
where \( \boldsymbol{\tilde{q}}_{u,g} \) is the composite eCSI vector across all transmit antennas, and \( \boldsymbol{\Lambda} = \text{Blkdiag}\{\boldsymbol{\alpha}_1, \dots, \boldsymbol{\alpha}_M\} \) is a block-diagonal matrix formed by the electromagnetic-domain precoders.

\begin{figure}[!t]
	%\vspace*{-5mm}
	\begin{center}
		\includegraphics[width = 1\columnwidth]{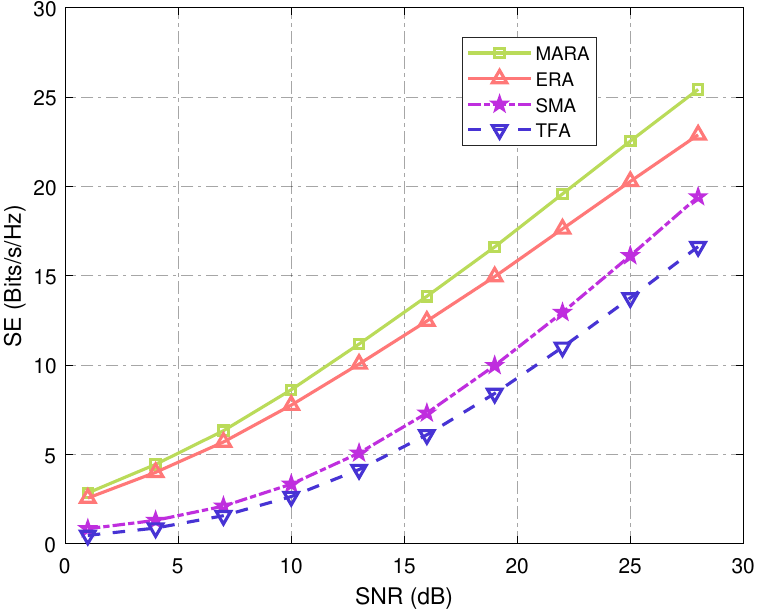}
	\end{center}
	%\vspace*{-5mm}
	% \captionsetup{font={footnotesize}, singlelinecheck=off, name={Fig.}, labelsep=period}
	\caption{The SE for MARA, ERA, SMA, and TFA.}
	\label{Fig.4antennaSE} % Fig.1
	\vspace*{-5mm}
\end{figure}

Owing to the normalization property inherent in spherical harmonic basis functions, the power constraint imposed on any radiation pattern \( f_{\text{tx},m}(\theta, \phi) \) becomes mathematically equivalent to enforcing \( \|\boldsymbol{\alpha}_m\|^2 = 1 \) on the corresponding electromagnetic-domain precoder. This fundamental equivalence can be rigorously demonstrated by incorporating equation (\ref{fdecomposition}) into the standard normalization condition \( \iint |f(\theta, \phi)|^2 \sin\theta d\theta d\phi = 1 \). Based on this critical insight, the SE maximization framework for ERA systems is established as follows
\begin{align}\label{SE2}
	\begin{split}
		&\mathop{\max}\limits_{\{\bm{\alpha}_m\}_{m=1}^{M},\left\{\bm{W}_{g}\right\}_{g=1}^{G}} R_{\rm ERA} \\ &
		\qquad\text{s.t.} \quad\|\bm{\alpha}_{m}\|^{2} = 1, \forall m, \\& \qquad\qquad\sum_{g=1}^{G}\|\bm{W}_g\|_{\mathrm{F}}^{2}\le P_T.
	\end{split}
\end{align}

\subsubsection{Channel Modeling and Optimization Problem of MARA}\label{S2.2.3}

Looking ahead to the development of future wireless systems, a single type of SMA or ERA may not fully meet all requirements. Therefore, hybrid antennas that jointly consider spatial and electromagnetic reconfigurability will become a trend. For instance, a BS antenna array can not only physically adjust its array shape (e.g., unfolding, folding, or integral translation/rotation) but also dynamically control the radiation pattern and amplitude-phase distribution of each array element. Such hybrid reconfigurable antennas combine the advantages of both SMA and ERA, on the one hand, they achieve large changes in channel gain through macroscopic displacement, on the other hand, they implement fine-grained beam and interference control via microscopic pattern reconfiguration.

The received signal at the \( g \)-th subcarrier of MARA is consistent with the channel models of SMA and ERA introduced earlier, where \( \boldsymbol{h}_{u,g}^H = [h_{u,1,g}^{\rm MARA}, h_{u,2,g}^{\rm MARA}, \dots, h_{u,M,g}^{\rm MARA}] \in \mathbb{C}^{1 \times M} \) signifies the downlink channel from the BS to the \( u \)-th UE. The only difference is that in this case, the channel position-related matrix \(\mathbf{B}_{u,m}\), the electromagnetic-domain precoders \(\boldsymbol{\alpha}_m\), and the digital precoders  \( \boldsymbol{w}_{u,g} \) need to be optimized simultaneously.

The channel representation can be simplified to the following compact formulation
\begin{equation}
	\label{split}
	h_{u,m,g}^{\rm MARA} = \boldsymbol{f}_{\text{rx},u}^T \boldsymbol{A}_u \boldsymbol{\Sigma}_{u,g} \boldsymbol{{B}}_{u,m} \boldsymbol{\Omega}_u \boldsymbol{\alpha}_m \stackrel{(b)}{=} \boldsymbol{{q}}_{u,m,g}^H \boldsymbol{\alpha}_m, 
\end{equation}
where $\bm{{q}}_{u,m,g}^{H} \triangleq \bm{f}_{{\rm{rx}},u}^{T}\bm{A}_{u}\bm{\mathit{\Sigma}}_{u,g}\bm{{B}}_{u,m}\bm{\mathit\Omega}_u\in \mathbb{C}^{1\times K}$.
Similarly, we can obtain the SE of the system as
\begin{equation}\label{R3}
	R_{\rm MARA}\! = \sum\limits_{g=1}^{G}\sum\limits_{u=1}^{U}\! \log_{2}\! \left(\!\! 1\! +\! \frac{|\bm{q}_{u,g}^{ H}{\bm{\mathit\Lambda}}\bm{w}_{u,g}|^2}{\sum\limits_{u^{\prime}\neq u}^{U}\! |\bm{q}_{u,g}^{H}{\bm{\mathit\Lambda}}\bm{w}_{u^{\prime},g}|^2\! +\! \sigma_n^2}\! \right)\!\! , \!
\end{equation}
where \( \boldsymbol{{q}}_{u,g} \) denotes the composite eCSI vector that now depends on the optimizable antenna positions through \(\boldsymbol{B}_{u,m}\).
Therefore, the SE optimization problem for MARA can be formulated as
\begin{align}\label{SE3}
	\begin{split}
		\mathop{\max}\limits_{\{\bm{p}_{m}\}_{m=1}^{M},\left\{\bm{\alpha}_m\}_{m=1}^{M},\{\bm{W}_{g}\right\}_{g=1}^{G}} R_{\rm MARA} \\ 
		\qquad\text{s.t.} \quad\|\bm{\alpha}_{m}\|^{2} = 1, \forall m, \\ \qquad\qquad\sum_{g=1}^{G}\|\bm{W}_g\|_{\mathrm{F}}^{2}\le P_T, \\
        \quad\bm{p}_{m}\in \mathcal{P}_m, \forall m.
	\end{split}
\end{align}

The optimization problem (\ref{SE3}) is highly non-convex due to the coupling among antenna positions \( \{\boldsymbol{p}_m\} \), electromagnetic-domain precoders \( \{\boldsymbol{\alpha}_m\} \), and digital precoders \( \{\boldsymbol{W}_g\} \). To solve this problem and validate the performance gains of the MARA paradigm, we employ a heuristic coordinate descent approach combined with inner-loop precoder optimization. Specifically, the antenna positions are sequentially optimized via exhaustive search over discretized candidate ports, where for each candidate position configuration, the electromagnetic-domain and digital precoders are jointly optimized using the WMMSE algorithm and manifold optimization, respectively. This iterative procedure continues until convergence. While this heuristic approach yields locally optimal solutions rather than global optima, it suffices to demonstrate the SE advantages of MARA over SMA, ERA, and conventional TFA architectures. More advanced AI-based solution approaches that can efficiently explore the vast configuration space are comprehensively reviewed in Section III.

Simulations were conducted for the above-mentioned scenarios under the following setup: The BS is equipped with a $4 \times 4$ uniform planar array (UPA) with $M = 16$ antennas, serving $U = 4$ single-antenna UEs. For SMA and MARA, each antenna can select from $3 \times 3 = 9$ candidate ports with inter-port spacing of $\lambda/5$. The system operates at carrier frequency $f_c = 3.5$ GHz with $G = 128$ OFDM subcarriers and subcarrier spacing $\Delta f = 30$ kHz. The number of spherical harmonic (SH) basis functions is $K = 10^2 = 100$. A multipath channel model with $L_u = 6$ paths per UE is adopted, with maximum delay spread $\tau_{\max} = 100$ ns. The azimuth angle of departure (AoD) for each path at the BS is assumed to be uniformly distributed within $\mathcal{U}[-60^{\circ}, 60^{\circ}]$, while the zenith AoD for each path follows $\mathcal{U}[60^{\circ}, 150^{\circ}]$. Perfect CSI is assumed at the BS. The average signal-to-noise ratio (SNR) is set to 15 dB. Results are averaged over 1000 independent Monte Carlo channel realizations.

The results are presented in Fig. \ref{Fig.4antennaSE}. It can be observed that the SE performance ranks in descending order as follows: MARA, ERA, SMA, and the conventional TFA. Quantitatively, compared with TFA, SMA achieves approximately 39\% SE improvement through spatial position optimization, ERA provides about 58\% gain via electromagnetic reconfiguration, while the proposed MARA paradigm attains a remarkable 107\% enhancement by jointly exploiting both spatial and electromagnetic DoFs. This indicates that the higher the degree of exploitation of the antenna's DoF, the higher the system's SE. Notably, the MARA gain exceeds the sum of individual SMA and ERA gains, suggesting a synergistic effect when both optimization dimensions are exploited simultaneously. These substantial performance gaps motivate the development of efficient AI-based optimization methods reviewed in Section III.

\subsection{Motivation for AI-Based Methods}\label{S2.3}

The optimization problems formulated in (\ref{SE1}), (\ref{SE2}), and (\ref{SE3}) present fundamental challenges that render traditional optimization methods inadequate: (i) \textit{non-convexity} arising from coupled position vectors and unit-norm constraints; (ii) \textit{infinite-dimensional search space} due to continuous antenna positions and exponentially growing electromagnetic parameter combinations scaling as $\mathcal{O}(K^M)$; (iii) \textit{real-time requirements} that iterative algorithms cannot meet within channel coherence time; and (iv) \textit{CSI acquisition bottleneck} where channel estimation depends on antenna configurations that themselves require CSI for optimization.

These challenges motivate AI-based methodologies reviewed in Section III, organized into three directions corresponding to MARA's core signal processing tasks:
\begin{itemize}
    \item \textbf{Channel Estimation (Section III-A):} Bayesian methods (S-BAR, SBL) \cite{successive,sbl}, supervised DL for joint sampling-estimation learning \cite{10925543}, and self-supervised architectures (AGMAE) exploiting spatial correlations via Transformer and graph attention \cite{AGMAE}.
    \item \textbf{Beamforming (Section III-B):} RL-based methods (DQN, PPO, MADDPG) for configuration space exploration \cite{lbj,bdf}, supervised DL (CNN, MLP) for fast CSI-to-configuration mapping \cite{dlf,dle}, and physics-informed hybrid approaches \cite{dla}.
    \item \textbf{ISAC (Section III-C):} Diffusion models for generative beam design \cite{isacnew2}, Transformer-based sequence models for adaptive alignment \cite{isacnew3}, physics-informed neural networks \cite{isacnew4}, and foundation models for cross-scenario adaptation \cite{isacnew5}.
\end{itemize}

\section{AI-Empowered Approaches for MARA}\label{S3}
\begin{figure}[!t]
	%\vspace*{-5mm}
	\begin{center}
		\includegraphics[width = 1\columnwidth]{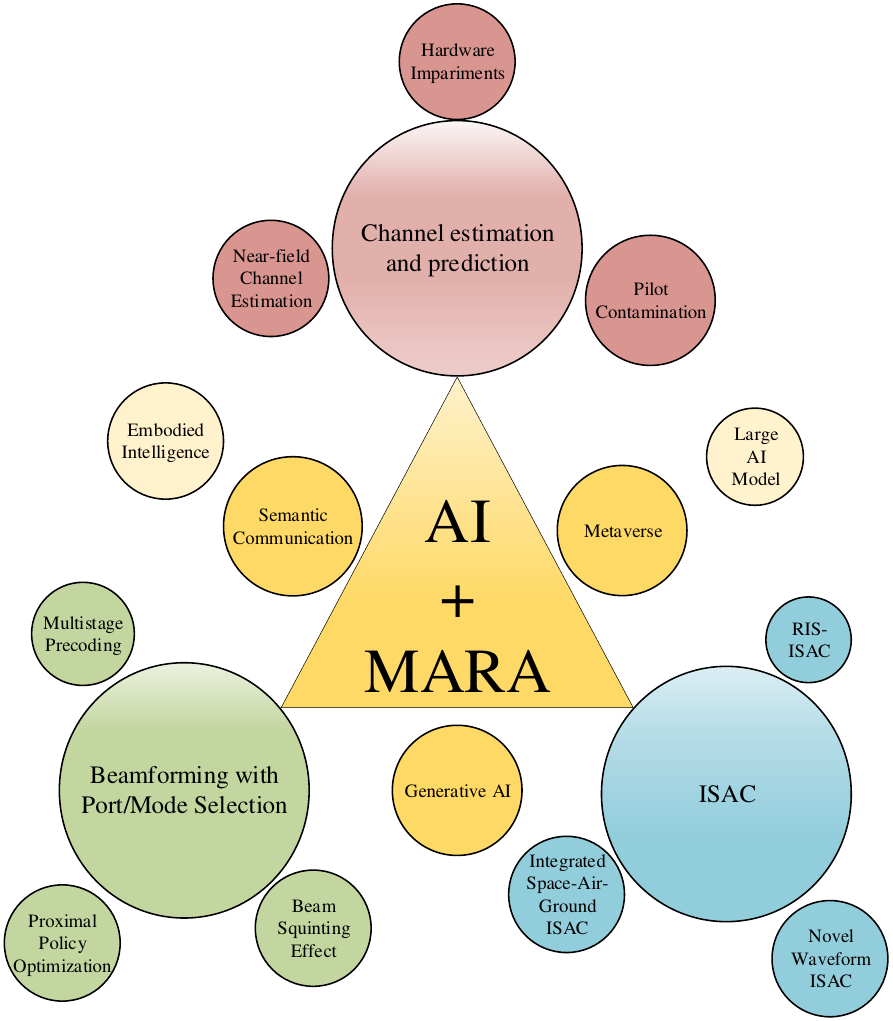}
	\end{center}
	%\vspace*{-5mm}
	% \captionsetup{font={footnotesize}, singlelinecheck=off, name={Fig.}, labelsep=period}
	\caption{Technical overview of ``AI+MARA".}
	\label{Fig.ai0} 
	\vspace*{-5mm}
\end{figure}

This chapter develops an AI-driven signal processing framework tailored to MARA’s key characteristics and associated optimization challenges. It covers three core research directions: first focusing on channel estimation and prediction, which leverages integrated sampling and learning approaches to derive accurate CSI from limited measurements, then addressing beamforming with port/mode selection, which adopts optimized designs under practical constraints to enhance system SE, fairness, and robustness, and finally exploring ISAC, which analyzes the tradeoffs introduced by MARA and proposes learning-based solutions to coordinate sensing and communication functions in dynamic scenarios. Fig. \ref{Fig.ai0} illustrates the chapter’s architecture.

\subsection{Channel Estimation and Prediction}\label{S3.1}

If it is desired to maintain the MARA system in an optimal working state, or to design beamforming and precoding, the system must first acquire the channel information for the entire movement area, i.e., the complete CSI.

However, acquiring CSI in a MARA system is far more challenging than in a TFA system, with the challenges stemming primarily from two aspects. First, it comes from the flexibility of the antenna's position and the requirement for high resolution. MARA systems have a large number of candidate antenna positions, which means the dimension of the channel to be estimated is extremely high \cite{buchong5}. If traditional methods are used to measure each antenna position, it will lead to unbearable pilot overhead, a task that is nearly impossible to complete within the channel's coherence time. Second, different radiation patterns may correspond to different channel states, which results in a rapid increase in the number of channel states \cite{10,14}.

Therefore, a more feasible approach is to estimate the CSI for only a small number of MARA positions and then use extrapolation techniques to predict the remaining positions. This chapter will detail an AI-driven signal processing paradigm specifically designed to address these challenges. We will systematically review how AI, through channel extrapolation techniques, can accurately recover the complete CSI from a small number of observed ports.

In MARA systems, which are characterized by physical movement, the antenna can move freely within a predefined continuous space, introducing a spatial dimension to channel estimation. In such conditions, we can reshape the channel estimation problem into a spatial optimization problem, i.e., intelligently determine where the antenna should move to measure, thereby obtaining the highest accuracy of CSI with the lowest overhead.
This is primarily divided into two technological approaches: intelligent sampling and reconstruction based on Bayesian machine learning \cite{successive,sbl}, and end-to-end joint optimization based on DL \cite{10925543,AGMAE}.

\begin{figure}[!t]
	%\vspace*{-5mm}
	\begin{center}
		\includegraphics[width = 1\columnwidth]{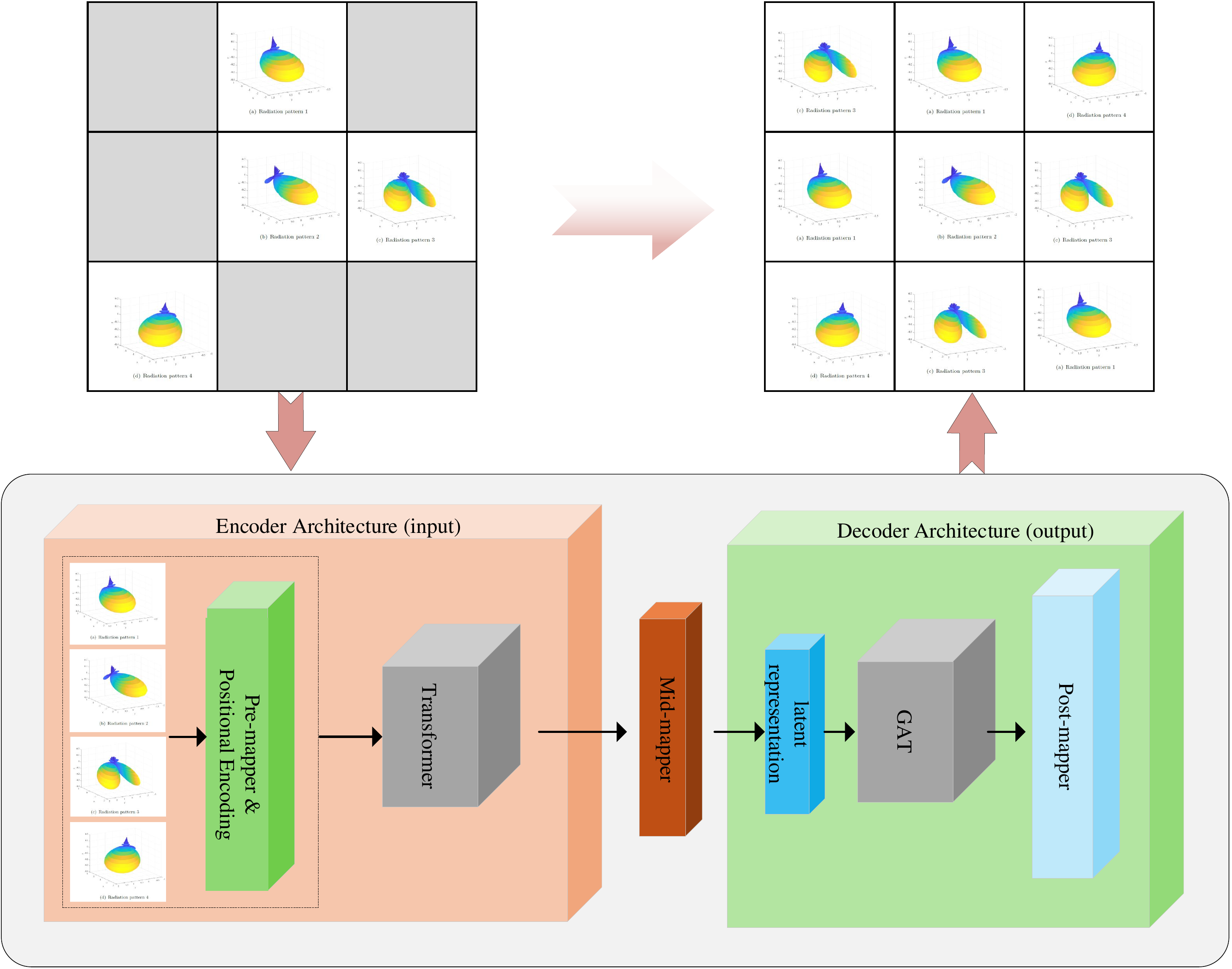}
	\end{center}
	%\vspace*{-5mm}
	% \captionsetup{font={footnotesize}, singlelinecheck=off, name={Fig.}, labelsep=period}
	\caption{System architecture of the AGMAE.}
	\label{Fig.ai1} 
	\vspace*{-5mm}
\end{figure}

Bayesian methods, an important branch of machine learning, handle uncertainty through probabilistic models, making them well-suited for MARA channel estimation. The core of Bayesian methods is to model the channel field across the entire movement area as a random process, and then to progressively reduce the uncertainty of the entire channel field by selecting sampling points.

Many traditional channel estimation algorithms rely on specific channel model assumptions, such as sparsity in the angular domain. When the actual channel environment does not conform to these assumptions, a model mismatch occurs, leading to a severe degradation in estimation performance \cite{successive}.
One approach is utilizing successive Bayesian reconstructor (S-BAR). S-BAR is a general estimation algorithm that does not rely on specific channel models \cite{successive}. It regards the channel estimation process as a sequential decision problem. Firstly, it models the channel as a Gauss process. Then, S-BAR adopts a greedy strategy, where at each step the location with the highest current uncertainty is selected for the next measurement. This is equal to guiding the antenna to move to the largest amount of information location to collect a sample. When a measurement is completed at a new location, S-BAR utilizes the Bayesian regression formula, incorporating the new measurement value to update the posterior mean and posterior variance of the entire channel field. This process is iterated until the uncertainty of the entire channel field is reduced below the threshold.	
On the other hand, in \cite{sbl}, they consider a spatially sparse clustered channel model. Sparse Bayesian learning (SBL) provides a powerful estimation framework. Due to the sparsity of the channel, SBL treats the sparse virtual channel vector to be solved as a random vector and applies Bayesian inference on this basis. The algorithm assigns a complex Gaussian prior distribution, controlled by an independent hyperparameter, to each basis and employs the Expectation-Maximization algorithm to estimate them. Eventually, the values of the hyperparameter converge, yielding the channel estimate.

The principles of spatial optimization can also be leveraged through DL. For instance, in \cite{10925543}, end-to-end joint learning of the measurement positions and the channel estimation function is implemented. 
It finds that the physical model for receiving pilot signals can be mathematically decomposed into an operational structure identical to that of a neural network layer. A weight matrix (representing the antenna position matrix) is multiplied by an input signal (representing the angle matrix), and then passed through a non-linear activation function (representing the complex exponential operation).
During the training stage, a deep neural network is constructed, whose first layer is specially designed to imitate the physical process described above. The weight matrix of this layer is constrained to represent the physical position coordinates of the antennas.
After training on a large amount of data, the weight matrix of the network's first layer converges to an optimal solution, which serves as the optimal set of antenna measurement positions. The rest of the network then constitutes an optimal channel angle estimation function.

In MARA systems characterized by electromagnetic reconfiguration, especially reconfigurable pixel antennas, control the current distribution on the antenna surface via electronic switches, thereby achieving flexible changes in the radiation pattern. This can be regarded as an equivalent form of antenna movement. This paradigm opens up new dimensions for channel optimization.
A DL method called the Asymmetric Graph Masked Auto-Encoder (AGMAE) was proposed in \cite{AGMAE}.
As shown in Fig. \ref{Fig.ai1}, AGMAE adopts an asymmetric encoder-decoder architecture. Its encoder, based on a Transformer, utilizes an attention mechanism to construct basis vectors from the known CSI. Meanwhile, its decoder, based on a graph attention network, recovers the CSI of unknown ports by leveraging the channel's local correlation and smoothness through a local diffusion mechanism. This is vital to capturing the spatial local correlation and smoothness of the channel across the antenna surface, thereby enabling more accurate reconstruction.

To provide a systematic overview of the AI methods discussed above for MARA channel estimation and prediction, Table \ref{tab:ce_taxonomy} presents a comprehensive taxonomy that classifies these approaches according to their methodological foundations, highlights their respective advantages and limitations, and identifies their most suitable application scenarios.

\begin{table*}[!t]
\centering
\caption{Taxonomy of AI Methods for MARA Channel Estimation and Prediction}
\label{tab:ce_taxonomy}

\begin{tabular}{|p{2.2cm}|p{2.8cm}|p{4cm}|p{4cm}|p{3cm}|}
\hline
\textbf{Category} & \textbf{Representative Methods} & \textbf{Key Advantages} & \textbf{Limitations} & \textbf{Application \newline Scenarios} \\
\hline
Bayesian Methods \cite{successive,sbl} & S-BAR, SBL& Model-agnostic; uncertainty quantification; no training data needed; sequential optimization & Computational complexity scales poorly with dimension; sequential nature limits parallelization & Low-mobility scenarios with limited pilot samples \\
\hline
Supervised DL \cite{10925543} & End-to-end DNN & Fast inference; jointly learns optimal sampling positions and estimation function & Requires large labeled datasets; limited generalization to new environments & Static environments with sufficient training data \\
\hline
Self-supervised DL \cite{AGMAE} & AGMAE & Exploits spatial correlations via attention; no labeled data required; captures local smoothness & Complex architecture; high training computational cost & Large-scale MARA with rich spatial structure \\
\hline
\end{tabular}
\end{table*}

The comparison in Table \ref{tab:ce_taxonomy} reveals a fundamental trade-off between model dependency and data efficiency. Bayesian methods such as S-BAR and SBL require minimal training data and provide principled uncertainty quantification, making them attractive for scenarios with limited pilot resources. However, their computational complexity becomes prohibitive for large-scale MARA systems. In contrast, DL-based methods achieve superior estimation accuracy and faster inference once trained, but they require substantial training data and may struggle with distribution shifts when deployed in new environments. The emergence of self-supervised approaches like AGMAE represents a promising middle ground, leveraging the spatial structure of MARA channels without requiring labeled optimal configurations.

\subsection{Beamforming with Position/Mode Selection}\label{S3.2}

In integrated sensing, communication, and power transfer (ISCPT) systems, the introduction of MARA provides a unified platform that can simultaneously satisfy user data rate demands, enable reliable target sensing, and deliver wireless power to energy-constrained devices \cite{alb,buchong7,buchong8}. By repositioning antennas, the effective aperture of the array is expanded, leading to finer angular resolution in sensing tasks and higher SNR in communication links. Furthermore, the joint optimization of beamforming vectors and antenna positions allows the system to lower the Cramér-Rao Bound (CRB) for target estimation accuracy while meeting strict communication and energy harvesting constraints. Similarly, in ISAC frameworks, MARA overcomes the rigidity of traditional uniform arrays, enhancing radar sensing precision while maintaining communication quality. Such dual-functionality makes MARA highly attractive in vehicular networks, smart cities, and industrial IoT scenarios where reliable connectivity and precise environment perception are equally critical.

From an optimization perspective, MARA-based systems introduce highly non-convex problems due to the coupling between antenna positions and beamforming weights. Conventional methods such as alternating optimization or convex relaxation can provide feasible solutions but often incur prohibitive computational costs. Analytical studies have derived closed-form solutions for certain special cases, for example, optimal receive MARA configurations that maximize sensing gain by symmetrically distributing antennas across the array aperture, achieving up to 4.77 dB improvements over TFA. For transmit-side optimization, boundary traversal search algorithms (BT-BFS and BT-DFS) and majorization-minimization approaches have been proposed to locate global or stationary solutions under line-of-sight (LoS) and non-line-of-sight (NLoS) environments. These results provide valuable theoretical insights into the performance upper bounds of MARA-enabled ISAC systems, but they remain limited in scalability and real-time adaptability when deployed in large-scale or fast-varying scenarios.

The growing complexity of dynamic wireless environments has motivated a shift toward machine learning-based solutions. RL has been particularly influential, as it can continuously adapt to environmental changes without requiring explicit mathematical models of the channel. Multi-agent deep reinforcement learning (DRL) frameworks such as heterogeneous MADDPG allocate dedicated agents to beamforming and antenna positioning tasks, enabling decentralized yet coordinated decision-making \cite{lbj}. These architectures are well-suited for large systems where centralized optimization becomes infeasible. In secure ISAC scenarios, proximal policy optimization (PPO)-based RL algorithms have been employed to jointly design beamforming and antenna movements under imperfect eavesdropper CSI, significantly improving secrecy rate robustness \cite{bdf}. Compared to static convex solvers, these learning-driven strategies achieve superior adaptability and can operate effectively in uncertain or adversarial conditions, bridging the gap between theoretical optimization and practical deployment.

\begin{figure}[!t]
	%\vspace*{-5mm}
	\begin{center}
		\includegraphics[width = 1\columnwidth]{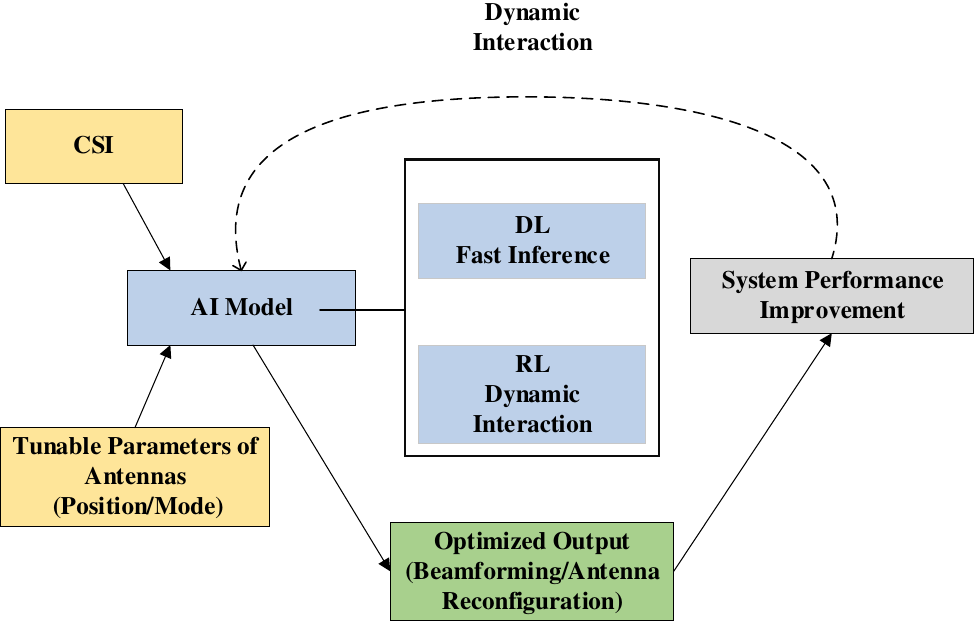}
	\end{center}
	%\vspace*{-5mm}
	% \captionsetup{font={footnotesize}, singlelinecheck=off, name={Fig.}, labelsep=period}
	\caption{An exemplary implementation of DRL for MARA.}
	\label{Fig.ai2} 
	\vspace*{-5mm}
\end{figure}

Parallel to RL, supervised and unsupervised DL approaches have achieved considerable success in beamforming and antenna position optimization, as shown in Fig. \ref{Fig.ai2}. CNN-based frameworks exploit the spatial structure of wireless channels to jointly infer optimal antenna configurations and beamforming vectors in two-dimensional MARA \cite{dlf}. In multicast scenarios, novel DL models composed of feature extraction, antenna position optimization, and beamforming modules have been developed, enabling end-to-end training with customized loss functions designed to maximize the minimum user gain \cite{dle}. Such methods achieve higher user fairness and robustness compared to iterative optimization algorithms, while drastically reducing computational complexity during inference. In anti-jamming contexts, multi-layer neural networks have been applied to jointly optimize beamforming and antenna displacement, producing near-optimal SINR performance with only marginal online computational costs \cite{dla}. These studies highlight DL’s ability to generalize across different environments once trained, making it a powerful tool for real-time MARA-enabled system design.

In terms of security and resilience, MARA provides a unique layer of physical protection against eavesdropping and malicious interference. By dynamically altering antenna geometry, MARA-enabled BS can degrade the channel quality of unintended receivers while enhancing legitimate user links. For example, DRL-assisted secure ISAC systems have demonstrated secrecy rate improvements of around 40\% under imperfect CSI conditions, substantially outperforming TFA systems \cite{bdf}. Similarly, MARA-aided anti-jamming designs leverage antenna displacement to spatially nullify jamming sources, enabling reliable communication even under multi-adversary scenarios \cite{dla}. The combination of mobility and adaptive learning thus introduces a flexible, physical-layer mechanism to ensure confidentiality and robustness, which is increasingly critical in open wireless environments such as satellite communication, UAV networks, and vehicular platoons.

The scope of MARA research is also expanding beyond linear and planar arrays toward higher-dimensional and conformal array structures. In 2D planar MISO systems, CNN-enabled flexible precoding strategies allow antennas to be repositioned along both horizontal and vertical axes, providing substantial gains in SE and adaptability to dynamic channels \cite{dlf}. More advanced architectures such as flexible cylindrical arrays (FCLA) extend antenna mobility to circular and vertical domains, allowing elements to revolve along circular tracks and shift between layers. Joint optimization via fractional programming and constrained adaptive moment estimation algorithms (CGS-Adam) has achieved up to 31\% throughput improvements in multi-user systems compared to fixed-position baselines \cite{fca}. These multidimensional approaches demonstrate the potential of MARA systems to achieve omnidirectional coverage, enhanced multiplexing, and robust interference management, paving the way for 6DMA that integrates both antenna spatial displacement and rotational DoF \cite{zhangrui2,zhangrui3,zhangrui4,zhangrui5}.

To systematically summarize the AI methods for MARA beamforming and position optimization discussed above, Table \ref{tab:bf_taxonomy} provides a comprehensive classification that compares different methodological approaches in terms of their optimization strategies, key strengths and weaknesses, and suitable deployment scenarios.

\begin{table*}[!t]
\centering
\caption{Taxonomy of AI Methods for MARA Beamforming and Position Optimization}
\label{tab:bf_taxonomy}

\begin{tabular}{|p{2.2cm}|p{2.8cm}|p{4cm}|p{4cm}|p{3cm}|}
\hline
\textbf{Category} & \textbf{Representative Methods} & \textbf{Key Advantages} & \textbf{Limitations} & \textbf{Application \newline Scenarios} \\
\hline
RL-based Methods \cite{lbj,bdf} & DQN, PPO, MADDPG & Adapts to dynamic environments; no explicit channel model required; handles continuous action spaces & Long training time; sample inefficiency; stability issues during training & Time-varying channels; security-critical applications; adversarial scenarios \\
\hline
Supervised DL \cite{dlf,dle} & CNN, MLP& Ultra-fast inference; end-to-end optimization; learns complex CSI-to-configuration mappings & Requires optimal labels from offline solvers; poor out-of-distribution generalization & Real-time systems with stable channel statistics \\
\hline
Hybrid Methods \cite{dla} & Physics-informed NN, Model-based DL & Improved generalization; physically interpretable solutions; reduced training data requirements & Complex architecture design; requires domain expertise & Safety-critical applications; limited training data regimes \\
\hline
\end{tabular}
\end{table*}

The taxonomy in Table \ref{tab:bf_taxonomy} highlights several important trends and trade-offs. RL-based methods have gained prominence due to their ability to learn adaptive policies without requiring labeled optimal solutions, making them particularly valuable for dynamic MARA scenarios where channel conditions change rapidly. However, their sample inefficiency and training instability remain challenges for practical deployment. Supervised DL methods offer the fastest inference times, suitable for real-time applications, but their reliance on pre-computed optimal labels limits their applicability in scenarios where such labels are expensive to obtain. The emerging hybrid approaches that integrate physical domain knowledge into neural architectures represent a promising direction, offering improved generalization and interpretability while maintaining computational efficiency. Overall, the field is evolving toward methods that can balance adaptability, computational efficiency, and generalization capability.

\subsection{ISAC}\label{S3.3}

The integration of sensing and communication functionalities in a MARA system introduces unique challenges due to fundamentally conflicting requirements between these two operations. High-precision sensing often demands wide coverage or agile beam scanning to detect and track targets, whereas communication typically requires focused, high-gain beams to maximize data rates to specific users. Satisfying both objectives simultaneously forces trade-offs in beam alignment, time-frequency resource allocation, and waveform design. Moreover, MARA’s dual reconfigurability (physical position adjustment and electromagnetic pattern control) creates strongly coupled design variables, dynamic antenna movements or pattern shifts intended to improve sensing performance can perturb the communication link (and vice versa), entangling the system’s geometry and channel conditions in a highly non-linear manner. This coupling, combined with the essentially infinite configuration space of continuous antenna positions and myriad reconfigurable parameters, makes joint optimization extremely complex and computationally intractable with classical methods. Practical hardware limitations further exacerbate these issues, mechanical repositioning and reconfiguration cannot occur instantaneously or without error, imposing constraints on how rapidly the system can switch between sensing and communication modes. Nevertheless, the potential performance benefits of integrating MARA into ISAC are substantial. For example, a recent study reported that dynamically optimizing antenna positions alongside beamforming in an ISAC BS yielded about a 60\% improvement in overall system performance compared to a conventional fixed-antenna baseline \cite{isacnew1}. This promising gain motivates the exploration of intelligent strategies to coordinate MARA’s dual capabilities for sensing and communication.

AI-driven strategies are increasingly being employed to tackle the joint sensing-communication optimization in MARA systems. One promising direction is generative modeling for adaptive beam design. Diffusion models have recently been applied to ISAC with remarkable results. By treating beamforming as a generative task, diffusion models can learn the underlying distribution of optimal beam patterns that serve both communication and sensing needs. For example, a diffusion-based ISAC scheme was used to reconstruct a target’s electromagnetic signature as a 3D point cloud, while jointly designing the transmit beams to meet both sensing accuracy and data rate constraints \cite{isacnew2}. This approach enabled high-fidelity sensing (reconstructing target shape and material properties) without violating a minimum communication throughput requirement. Another cutting-edge approach leverages transformer-based sequence models to coordinate MARA’s reconfigurable operations over time. Transformers have a unique ability to capture long-range spatio-temporal dependencies through self-attention, making them well-suited for dynamic beam alignment and tracking problems. In a recent design, a causal transformer was trained to process sequences of pilot signals and past beamforming actions, yielding an adaptive beam alignment policy that outperformed recurrent neural network baselines across diverse channel conditions \cite{isacnew3}. The transformer’s inductive bias enables superior generalization in heterogeneous environments. These generative and deep sequence-learning techniques illustrate how AI can exploit MARA’s extra DoF. By creatively searching through the vast configuration space, they identify non-intuitive antenna positioning and beamforming strategies that jointly enhance sensing and communication performance beyond classical limits.

In tandem with purely data-driven methods, researchers are also exploring hybrid AI frameworks that embed physical domain knowledge into the learning process. Physics-informed neural networks (PINNs) and other model-based learning approaches integrate electromagnetic propagation laws or spatial constraints into neural architectures, thereby reducing the solution space and improving reliability of the learned policies. For instance, incorporating Maxwell’s equations or array manifold models as a regularization in DL can guide the MARA configuration toward physically plausible solutions, ensuring that the dual-use beam patterns adhere to known radar and communication theory \cite{isacnew4}. Likewise, large-scale pre-trained models are being developed to handle cross-domain tasks in ISAC. The idea is to train a unified model on a broad distribution of sensing and communication data so that it can adapt to new scenarios with minimal fine-tuning. A recent study demonstrated a foundation model for wireless channels that, after massive self-supervised pre-training on diverse channel data, could predict channel states for different network configurations without retraining \cite{isacnew5}. Extending such paradigms to ISAC hints at a future where one large model could seamlessly handle radar sensing and communication waveform design together, leveraging shared representations between the two domains. 

These examples demonstrate the powerful capabilities of AI in ISAC+MARA, particularly in terms of optimizing antenna configurations. AI can efficiently handle large-scale, nonlinear optimization problems through intelligent decision-making, which traditional analytical methods and solution strategies are often difficult to address. These AI methods can significantly improve system performance, making ISAC systems more adaptable, flexible, and efficient. In summary, AI technologies will bring significant gains to ISAC+MARA, and are likely to be a game-changer, becoming a core technology for the new generation of communications.

\section{Future Directions for MARA and Promising Roles of AI}

\subsection{Synergy of MARA and Novel Waveform}

6G demands waveforms beyond OFDM to handle mobility and frequency selectivity. Orthogonal time frequency space (OTFS) and MIMO-Chirp are strong candidates, leveraging delay-Doppler processing and spread-spectrum, respectively \cite{tutorial2.1}. Coupling them with MARA unlocks spatial diversity and adaptive propagation for joint performance gains. For OTFS, MARA’s real-time repositioning complements delay-Doppler symbol mapping, reducing path loss and boosting SINR. Recent results show higher rates for MARA-OTFS versus static arrays in high-mobility settings \cite{tutorial2.2}. In MIMO-Chirp systems, optimizing MARA spacing and orientation enhances Doppler robustness, cuts ICI, and improves range resolution for autonomous driving and industrial IoT. Beyond these, affine frequency division multiplexing (AFDM) has gained attention as a DAFT-based multicarrier that maintains orthogonality under high and fractional Doppler, offers strong delay-Doppler diversity on doubly selective channels, and exhibits favorable ICI behavior, which makes it a natural fit for MARA-assisted high-mobility links \cite{AFDM_NG}.

AI supports this synergy end-to-end, learned surrogates can map waveform and channel descriptors to near-optimal MARA poses and electromagnetic patterns, policy learning can co-adapt chirp parameters with spatial moves to stabilize delay-Doppler sparsity and suppress ICI, and generative co-design can propose waveform-MARA configurations tailored to scene geometry \cite{tutorial2.3}. At mmWave/THz, where path loss and absorption are severe, MARA’s user tracking and beam steering complement the Doppler resilience of OTFS and AFDM, preserving throughput while widening unambiguous Doppler and improving sensing ambiguity characteristics. Looking ahead, federated learning can coordinate online adaptation across MARA nodes with limited pilots, while AFDM’s low-dimensional DAFT parameterization facilitates fast retuning under tight latency and energy budgets \cite{AFDM_NG}.

\subsection{The Combination of MARA and Embodied Intelligence}

The convergence of MARA and embodied intelligence (EI) shifts antennas from passive transceivers to active elements of intelligent systems. EI’s tight coupling of body and environment aligns with MARA’s spatial/electromagnetic reconfigurability, enabling drones and robots to adapt communication and sensing to situational changes. In embodied systems, MARA functions as both link interface and environmental sensor. By exploiting reflections and channel fluctuations, MARA-equipped agents infer obstacle locations and material properties to enhance situational awareness \cite{tutorial4.1}. AI fuses MARA’s spatial cues with other modalities to form world models, while RL-style parameter tuning provides a basis for multimodal perception pipelines.

MARA’s reconfigurability sustains robust connectivity in dynamic settings (e.g., swarming UAVs, autonomous vehicles) by continuously adjusting positions and orientations, which is crucial for low-latency edge computing \cite{tutorial4.2}. Online learning can further adapt MARA to agent trajectories and traffic demands. A key constraint is energy, limited onboard power caps reconfiguration frequency and span \cite{tutorial4.3}. Promising remedies include energy-harvesting MARA designs, lightweight AI inference, and distributed optimization that coordinates multiple agents to balance system-level throughput with local sensing needs. Future directions include ``neuromorphic antennas” whose MARA states evolve with stimuli, enabling self-healing links in adverse conditions, and integration with digital twins for virtual co-design and stress-testing prior to deployment.

\subsection{Multi-stage Beamforming with Emerging Architectures}

\begin{figure}[!t]
	\begin{center}
		\includegraphics[width = 1\columnwidth]{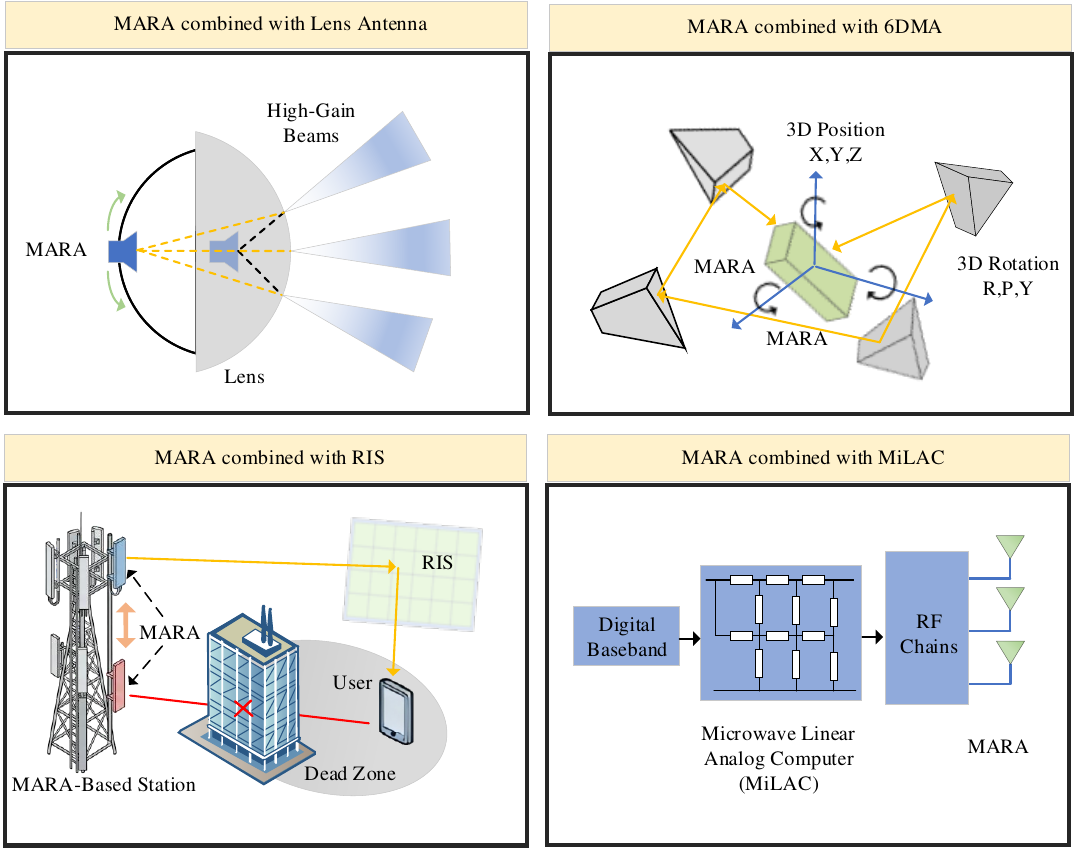}
	\end{center}
	\caption{Four representative multi-stage beamforming paradigms integrating MARA with emerging architectures: (a) Lens antenna integration; (b) 6DMA integration; (c) RIS integration; (d) MiLAC integration.}
	\label{Fig.multistage} 
	\vspace*{-5mm}
\end{figure}

The MARA paradigm activates new DoF through flexible spatial repositioning, yet this represents only a fraction of its full potential. By integrating MARA with emerging architectures such as lens antennas, 6DMA, RIS, and microwave linear analog computing (MiLAC), multi-stage beamforming systems can be constructed that achieve significant breakthroughs in energy efficiency, coverage, and computational complexity \cite{tutorial4.3}. Fig. \ref{Fig.multistage} illustrates four representative integration paradigms, each exploiting complementary physical mechanisms to enhance overall system performance.

\textbf{Integration with Lens Antennas:} As shown in Fig. \ref{Fig.multistage}(a), lens antennas employ dielectric lenses or metasurface domes to focus electromagnetic waves, substantially reducing the number of RF chains required for large-scale arrays. The combination of MARA with lens antennas enables the feed elements or arrays to move along the focal plane of the lens, thereby flexibly steering high-gain beam directions while maintaining high gain. This approach achieves wide-angle coverage and energy-efficient transmission at significantly reduced hardware costs \cite{lens_array_2024, wan2020compressive_lens}.

\textbf{Integration with 6DMA:} As illustrated in Fig. \ref{Fig.multistage}(b), 6DMA extends antenna mobility from three-dimensional position (3D position) to three-dimensional rotation (3D rotation), providing a complete six-dimensional spatial DoF. The fusion of MARA with 6DMA enables simultaneous optimization of antenna positions and orientations, maximally exploiting the spatial variations of multipath channels. This yields performance gains far exceeding those of conventional fixed-antenna systems in interference suppression, channel hardening, and physical-layer security \cite{zhangrui1}.

\textbf{Integration with RIS:} As depicted in Fig. \ref{Fig.multistage}(c), the combination of MARA (active-side mobility) with RIS (environment-side reconfiguration) forms an ``active-passive'' cooperative beamforming architecture. MARA can dynamically adjust positions to establish optimal LoS links with the RIS or circumvent obstacles to maximize the reflection gain provided by RIS. This synergistic effect demonstrates significant advantages in coverage extension for communication dead zones, simultaneous wireless information and power transfer (SWIPT), and ISAC scenarios \cite{mara_ris_swipt_2025}.

\textbf{Integration with MiLAC:} As shown in Fig. \ref{Fig.multistage}(d), MiLAC is an analog computing technology based on microwave network architectures that can perform complex operations such as matrix inversion and beamforming at the speed of light during electromagnetic wave propagation. Incorporating MiLAC into MARA systems enables ``instantaneous computation'' in the analog domain to replace energy-intensive digital baseband processing. This not only dramatically reduces the computational complexity and latency associated with processing massive MIMO signals, but also provides a novel hardware-based solution paradigm for achieving green communications in ultra-large-scale antenna arrays \cite{milac_part1_2025,milac_part2_2025}.

These multi-stage beamforming architectures collectively represent a paradigm shift from isolated antenna optimization toward holistic system co-design, where the transceiver, propagation medium, and signal processing are jointly optimized as a unified platform. AI-driven optimization techniques, particularly graph neural networks and physics-informed learning, can efficiently navigate the joint design space by exploiting the underlying network structures and electromagnetic constraints inherent in these emerging architectures.

% In conclusion, the future of MARA technology is intricately linked to advancements in AI and 6G communication. From synergistic integration with lenses and novel waveforms to enabling OTH ISAC and embodied intelligence, MARA's dynamic reconfigurability provides a foundation for next-generation wireless systems. Addressing the challenges of energy efficiency, real-time optimization, and multi-system integration will be critical to realizing this vision, with AI serving as the key enabler for intelligent adaptation and resource management.

\section{Conclusions}

This paper overviews the fundamentals and advancements of MARA-aided wireless networks. We first highlight MARA's motivation by analyzing TFA's limitations in dynamic 6G scenarios, then classify MARA into SMA, ERA, and their integration, clarifying their principles and trade-offs, and build a unified mathematical framework for wideband TDD-mMIMO-OFDM channel modeling and SE optimization to capture MARA-induced dynamic channel changes, with simulations verifying MARA's superiority over TFA, SMA, or ERA alone. Next, we review state-of-the-art AI-driven solutions for MARA's core challenges, including Bayesian/DL for CSI acquisition from limited pilots, RF/CNN-based methods for non-convex beamforming and port/mode selection, and meta-reinforcement learning for balancing communication-sensing performance in ISAC, along with prototype and experimental results verifying MARA's practical effectiveness. Finally, we identify promising MARA directions: synergy with novel waveforms for high mobility, combination with embodied intelligence for autonomous systems, and multi-stage beamforming with emerging architectures including lens antennas, 6DMA, RIS, and MiLAC for holistic system co-design. Given MARA research is in its infancy, gaps exist in hybrid reconfiguration modeling and AI framework implementation, and we hope this work guides researchers and engineers to unlock MARA's potential for adaptive, efficient 6G networks.

\bibliographystyle{IEEEtran}
\bibliography{refs}
\balance

\end{document}